\documentclass[preprint,aps,eqsecnum,showpacs,tightenlines]{revtex4}
\usepackage{epsfig}

\newcommand{\intk}{\int\!\frac{{d^3}k }{(2\pi)^3}\,}
\newcommand{\inty}{\int\!d^3 y\,}
\newcommand{\intx}{\int\!d^3 x\,}
\newcommand{\intxy}{\int\!d^3 x\, \int\! d^3 y\,}

\newcommand{\be}{\begin{equation}}
\newcommand{\ee}{\end{equation}}
\newcommand{\bea}{\begin{eqnarray}}
\newcommand{\eea}{\end{eqnarray}}
\newcommand{\kma}{\; ,}
\newcommand{\pkt}{\; .}
\newcommand{\call}{{\cal L}}
\newcommand{\calm}{{\cal M}}
\newcommand{\calf}{{\cal F}}
\newcommand{\calr}{{\cal R}}
\newcommand{\cale}{{\cal E}}
\newcommand{\inttt}{\int\limits_{0}^{t}\!{d}t'\,}

\newcommand{\oma}{\omega_{a0}}

\newcommand{\omel}{\omega_{e\lambda 0}}
\newcommand{\re}{{\rm Re}}
\begin{document}
\preprint{DO-TH 01/02, LA-01-356}
\title{\Large \bf Gauge Fields Out-Of-Equilibrium: A Gauge Invariant Formulation and the 
Coulomb Gauge }
\author{{\bf Katrin Heitmann}
\footnote{\noindent on leave of absence from Dortmund University\\ e-mail:~heitmann@lanl.gov} } 
\affiliation{T8, Theoretical Division, Los Alamos National Laboratory, Los Alamos, New Mexico 87545, U.S.A.}
\date{January 24, 2001}
\begin{abstract}
We study the abelian Higgs model out-of-equilibrium in two different
approaches, a gauge invariant formulation, proposed by Boyanovsky
et al. \cite{Boyanovsky:1996dc} and in the Coulomb gauge. We show
that both approaches become equivalent in a consistent one loop
approximation. Furthermore, we carry out a proper renormalization
for the model in order to prepare the equations for a numerical
implementation. The additional degrees of freedom, which arise
in gauge theories, influence the behavior of the system dramatically.
A comparison with results in the 't Hooft-Feynman background
gauge found by us recently, shows very good agreement.
\end{abstract}
\maketitle

\section{Introduction}
Non-equilibrium dynamics have become a very active field of research
in the last few years in nearly all parts of physics. 
In condensed matter physics for example the description of the dynamics
of non-equilibrium phase transitions plays an important role \cite{Zurek:1996sj}.
Such phase transitions occur in ferromagnets, superfluids, and liquid crystals
to name only a few. They are subjects of intensive studies, 
both theoretical and experimental. 

Also in cosmology  some phenomena require the use of
non-equilibrium technics. One example 
is the electroweak phase transition. The electroweak
phase transition could lead
to a possible model for baryogenesis.
This problem is investigated by
several groups, e.g., \cite{Garcia-Bellido:1999sv,Krauss:1999ng},
using non-equilibrium methods.

Another phenomenon in cosmology where non-equilibrium dynamics 
are important is the inflationary
phase of the early universe which is
studied intensively by different groups
\cite{Khlebnikov:1998sz,Linde:1998gs,Greene:1997fu,Kofman:1997yn,Ramsey:1997sa,Ramsey:1998fz,Greene:1998nh,Boyanovsky:1996sq,Boyanovsky:1997cr,Baacke:1999gc}. Inflation refers
to an epoch during the evolution of the universe in which it underwent
an accelerated expansion phase. This would resolve some of the short comings of the 
Standard hot Big Bang model, e.g., the flatness problem, concerning the
energy density of the universe and the horizon problem, related by the large
scale smoothness of the universe, indicated by the Cosmic Microwave Background
Radiation (CMBR). 

At lower energies heavy ion collisions are under consideration as non-equilibrium
processes \cite{Boyanovsky:1993vi,Boyanovsky:1995yk,Kluger:1991ib,Cooper:1993hw,Cooper:1994hr,Cooper:1995ji}. In such heavy ion collisions a new state
of matter could be reached if the short range impulsive forces between
nucleons could be overcome and if squeezed nucleons would merge into each other.
This new state should be a Quark Gluon Plasma (QGP), in which quarks and gluons,
the fundamental constituents of matter, are no longer confined, but free
to move around over a volume in which a high enough temperature and/or density
reveals. Heavy ion collisions
 are studied experimentally at current and
forthcoming accelerators, the Relativistic Heavy Ion Collider RHIC at Brookhaven and
the Large Hadron Collider LHC at CERN. The occurring Quantum Chromodynamic (QCD) 
phase transition in these processes
could be out of equilibrium and lead to formations of coherent  
condensates of low energy pions,
so called Disoriented Chiral Condensates (DCC). Recent results reported from
CERN-SPS \cite{Heinz:2000bk} seem to indicate a strong evidence for the
existence of a QGP in Pb-Pb collisions.

The underlying theories of the above described phenomena are gauge theories.
Therefore, it is of great importance to implement gauge field fluctuations
if one wants to deal with more realistic models. 
Gauge theories out of equilibrium are still poorly explored in the literature.
While the aspects of various gauges and of gauge invariance have been
discussed extensively in perturbation theory and for equilibrium systems,
the analysis of such systems out of equilibrium is still to be developed.
The analysis presented here focus on different approaches for the description
of non-equilibrium gauge theories and compare these approaches.

In \cite{Baacke:1997kj} we have presented the theoretical framework and its numerical
implementation for the SU(2) Higgs model in the 't Hooft-Feynman background gauge.
Obviously, the question of gauge dependence and independence has to be explored.
In \cite{Baacke:1999sc} we investigated the gauge invariance of the one-loop effective
action of a Higgs field in the SU(2)-Higgs model for a time independent
problem, the bubble nucleation. An analogous analysis for a system out of equilibrium
\cite{Heitmann:2000} has shown that the fluctuation operator decomposes
in a similar way to the one derived in \cite{Baacke:1999sc}.  
We have  found within this analysis a close relation between the 't Hooft-background gauge
and the Coulomb gauge.

In this work we are now considering two other approaches in order to investigate
gauge theories out of equilibrium and to get some insight how the behavior
of the system is influenced by the choice of gauge. Therefore, we study
the Abelian Higgs model in a gauge invariant approach which was developed
by Boyanovsky et al. \cite{Boyanovsky:1996dc,Boyanovsky:1998dj} in order to
study non-equilibrium aspects of the effective potential.
We use their formalism in order to derive the equations of motion
in the one loop approximation. 
We will indicate that the gauge invariant approach leads to IR-problems
beyond the one loop approximation. As a second approach we
examine the Coulomb gauge. Here we use a gauge condition in order to eliminate
the unphysical degrees of freedom. We show how a proper renormalization
can be implemented. Therefore, we use a scheme which was developed by us for the $\phi^4$~theory
\cite{Baacke:1997se} and extended to fermionic and gauge 
systems \cite{Baacke:1998di,Baacke:1997kj}. Furthermore, we show how the gauge invariant 
approach and the Coulomb gauge are related to each other.  

In order to compare the behavior of the system for different gauges quantitatively,
we reexamine the 't Hooft-Feynman gauge and implement the 't Hooft-Feynman gauge
and the Coulomb gauge numerically. We examine the development of the
zero mode and the different fluctuations. We also investigate the problem of
energy conservation.

The paper is organized as follows. In section 2 we start our discussion by 
outlining the main aspects of the gauge invariant approach of
Boyanovsky et al. We derive the equation of motion for the zero mode and
the different fluctuations and we determine the energy. 
In section 3 we present the equation of motion and the energy
in the Coulomb gauge. The comparison of the two approaches is
investigated in section 4. In section 5 we show how the
renormalization can be done properly in order to prepare the equations
for numerical implementations. 
Some numerical results for the Coulomb gauge
and for comparison also in the 't Hooft-Feynman background gauge
are presented in section 5.
We conclude in section 6 with a discussion of the 
influence of the different degrees of freedom and an
outlook to more realistic applications of the method.

\section{A Gauge Invariant Approach}
\subsection{The Formalism}
\label{formalism}
We  study the Abelian Higgs model in a gauge invariant formulation.
The basic ideas for this description are developed in 
\cite{Boyanovsky:1996dc,Boyanovsky:1998dj}.
We give here a short overview about the derivation of the Hamiltonian.
For details the reader is referred to the two papers mentioned above.
The Lagrangian density for the Abelian Higgs model reads
\bea\label{lagrange}
{\cal L}&=&-\frac 1 4 F^{\mu\nu} F_{\mu\nu}+D_\mu\phi^\dagger D^\mu \phi
-\frac{\lambda}{4}(\phi^\dagger\phi-v^2)^2\kma\\
D_\mu&=&\partial_\mu+ie A_\mu\pkt
\eea 
We want to formulate a gauge invariant Hamiltonian. Therefore, we use
as in \cite{Boyanovsky:1996dc,Boyanovsky:1998dj} the canonical formulation.
We have to identify the canonical field variables and constrains. The
canonical momenta conjugate to the scalar and vector fields are given by
\bea
\Pi^0&=&0\kma\\
\Pi^i&=&\dot A^i+\nabla^i A^0=-E^i\kma\\
\pi^\dagger&=&\dot\phi+ieA^0\phi\kma\\
\pi&=&\dot\phi^\dagger-ieA^0\phi^\dagger\pkt
\eea
Therefore, the Hamiltonian reads 
\bea
H&=&\intx\left\{\frac 1 2 \vec \Pi\cdot\vec \Pi^{\dagger}
+\pi^\dagger\pi+(\vec\nabla\phi-ie\vec A\phi)\cdot(\vec\nabla\phi^\dagger
+ie\vec A\phi^\dagger)\nonumber\right.\\
&&+\frac 1 2 (\vec\nabla\times\vec A)^2
+\left.\frac{\lambda}{4}(\phi^\dagger\phi-v^2)^2
+A_0\left[\vec\nabla\cdot\vec \Pi-ie(\pi\phi-\pi^\dagger\phi^\dagger)
\right]\right\}\pkt
\eea
We will quantize this system with Dirac's method \cite{Henneaux:1992}. 
Therefore, we have to recognize the first class constraints (mutually
vanishing Poisson brackets). Then the constraints become operators
in the quantum theory and are imposed onto the physical states,
thus defining the physical subspace of the Hilbert space and gauge
invariant operators. We have two first class constrains
\be
\Pi^0=\frac{\delta{\cal L}}{\delta A^0}=0\kma
\ee
and Gauss' law
\bea
{\cal G}(\vec x,t)&=&\nabla^i\pi^i-\rho=0\kma\\
\rho&=&ie\left(\phi\pi-\phi^\dagger\pi^\dagger\right)\kma
\eea
with $\rho$ being the matter field charge density.
We can now quantize the system by imposing the canonical
equal time commutation relations
\bea
&&[\Pi^0(\vec x,t),A^0(\vec y,t)]=-i\delta (\vec x-\vec y)\kma\\
&&[\Pi^i(\vec x,t),A^j(\vec y,t)]=-i\delta^{ij}\delta (\vec x-\vec y)\kma\\
&&[\pi^\dagger(\vec x,t),\phi^\dagger(\vec y,t)]=-i\delta (\vec x-\vec y)\kma\\
&&[\pi(\vec x,t),\phi(\vec y,t)]=-i\delta (\vec x-\vec y)\pkt
\eea
In Dirac's formulation, physical operators are those that commute with
the first class constraints. Since $\Pi^0(\vec x,t)$ and ${\cal G}(\vec x,t)$
are generators of local gauge transformations, operators that commute
with the first class constraints are gauge invariant 
\cite{Boyanovsky:1996dc,Boyanovsky:1998dj}. As shown in 
\cite{Boyanovsky:1996dc}  the fields and the
conjugate momenta can be written in the following form
\bea
\Phi(\vec x)&=&\phi(\vec x)\exp\left[ie\inty \vec A(\vec y)\cdot\vec\nabla_y 
G(\vec y-\vec x)\right]\kma\\
\Pi(\vec x)&=&\pi(\vec x)\exp\left[-ie^2\inty\vec A(\vec y)\cdot\vec
\nabla_y G(\vec y-\vec x)\right]\kma
\eea
with $G(\vec y-\vec x)$ the Coulomb Green's function that satisfies
\be
\triangle G(\vec y-\vec x)=\delta^3 (\vec y-\vec x)\pkt
\ee
They are invariant under gauge transformations \cite{Boyanovsky:1996dc}.
The gauge field can be separated into transverse and longitudinal
components
\bea
&&\vec A(\vec x)=\vec A_L(\vec x)+\vec A_T(\vec x)\kma\\
&&\vec \nabla\times\vec A_L(\vec x)=0\kma\\
&&\vec\nabla\cdot\vec A_T(\vec x)=0\pkt\\
\eea
Since the fields $\vec A_T$ and $\Phi$ and their canonical momenta
commute with the constraints, they are gauge invariant. It is also
possible to write the momentum canonical to the gauge field in
longitudinal and transverse components
\be
\vec \Pi(\vec x)=\vec\Pi_L(\vec x)+\vec \Pi_T(\vec x)\kma
\ee
where both components are gauge invariant. In \cite{Boyanovsky:1996dc},
it is mentioned that in all matrix elements between gauge invariant
states the longitudinal component can be replaced by the charge density
\be
\vec \Pi_L(\vec x)\rightarrow ie\left[\Phi(\vec y)\Pi(\vec y)
-\Phi^\dagger(\vec y)\Pi^\dagger(\vec y)\right]=\rho\pkt
\ee
Finally, the Hamiltonian becomes
\bea
H&=&\intx \left\{\frac 1 2 \vec \Pi_T\cdot\vec \Pi_T
+\Pi^\dagger\Pi+(\vec\nabla\Phi-ie\vec A_T\Phi)\cdot
(\vec\nabla\Phi^\dagger+ie\vec A_T\Phi^\dagger)\right.
+\frac 1 2 (\vec\nabla\times\vec A_T)^2\nonumber\\
&&+\frac\lambda 4(\Phi^\dagger\Phi-v^2)^2\left.
+\frac 1 2 \intxy\rho(\vec x)G(\vec x-\vec y)\rho(\vec y)\right\}\pkt
\eea
The features of this Hamiltonian are discussed at length in 
\cite{Boyanovsky:1996dc,Boyanovsky:1998dj}. One of the striking points is
the equivalence with the Hamiltonian in the Coulomb gauge. This similarity 
is not uncommon because, in the Coulomb gauge, only the physical
degrees of freedom are taken into account. 
We  focus our interest on the non-equilibrium aspects of the 
theory.
In \cite{Boyanovsky:1996dc,Boyanovsky:1998dj}, the effective
potential was derived and some aspects of non-equilibrium dynamics
were discussed. But as there explain the effective potential
is not suitable for non-equilibrium dynamics. 
We derive here the full non-equilibrium equations.
We include not only the terms which are quadratic in the
zero mode, but all terms up to second order which are
relevant for the one loop approximation. We consider the 
derivative terms of the zero mode and of its conjugate momentum.
This yields the wave function renormalization which is not
considered in \cite{Boyanovsky:1996dc,Boyanovsky:1998dj}
or has to be introduced by hand. We will also see that the formalism
does not give a clear statement about the loop order which is included.
When linearized, the equations become equivalent to those obtaining in the Coulomb gauge,
which we consider in detail in section \ref{equiv}. There,
we show the correspondence of the Hamiltonian approach and
the Coulomb gauge. As we will show, the inclusion of higher loop terms in the Hamiltonian
approach leads to problems in the IR-region. 

First of all, we 
derive the  equation of motion for the fields.
Therefore, we separate the expectation value of $\Phi$ and of its
canonical momentum into
a mean value and a fluctuation part
\bea
\Phi(\vec x,t)&=&\phi(t)+\varphi(\vec x,t)\kma\\
\Pi(\vec x,t)&=&\Pi(t)+\pi(\vec x,t)\pkt
\eea
We also introduce real fields and canonical momenta as follows
\bea
\varphi&=&\frac{1}{\sqrt 2}(h+i\varphi)\kma\\
\pi&=&\frac{1}{\sqrt 2}(\pi_h-i\pi_\varphi)\pkt
\eea
Therefore, we find as a gauge invariant Hamiltonian
\bea
H&=&\Omega\left[\frac 1 2 \Pi^2+U(\phi)\right]\\
&&+\intx\left[\frac 1 2 \vec\pi_\bot^2-\frac 1 2 
(\vec\nabla\times\vec a_\bot)^2+\frac 1 2 e^2a_\bot^2\phi^2+\frac 1 2 \pi_h^2
+\frac 1 2 \pi_\varphi^2\right.\nonumber\\
&&\left.\hspace{1.5cm}
+\frac 1 2 (\vec\nabla h)^2+\frac 1 2 (\vec\nabla\varphi)^2
+\frac \lambda 2 (\phi^2-v^2)(h^2+\varphi^2)
+\lambda\phi^2 h^2
\right]\nonumber\\
&&+\intxy\frac {e^2}{2} \left[\phi\pi_\varphi(\vec x)-
\Pi\varphi(\vec x)\right]G(\vec x-\vec y)\left[
\phi\pi_\varphi(\vec y)-\Pi)\varphi(\vec y)\right]\nonumber\kma
\eea
with
\begin{equation}
U(\phi)=\frac \lambda4 (\phi^2-v^2)^2\pkt
\end{equation}
\subsection{Equation of Motion and Energy Density}
We are now able to derive the equations of motion for the zero mode and 
the fluctuation fields. In the Hamiltonian formalism we find 
the equation of motion by calculating the 
commutator between the Hamiltonian and the field.
We obtain for the zero mode $\phi$ 
\bea
\dot\phi
&=&\Pi\left[1+\frac{e^2}{\Omega}\intxy\varphi(\vec x)G(\vec x-\vec y)
\varphi(\vec y)\right]\\
&&-\frac{e^2}{2\Omega}\phi\intxy\left[
\varphi(\vec x) G(\vec x-\vec y)\pi_\varphi(\vec y)+\pi_\varphi(\vec x) G(\vec x-\vec y)
\varphi(\vec y)\right]\nonumber\kma
\eea
and for the canonical momentum
\bea
\dot\Pi
&=&-U'(\phi)-\frac{\phi}{\Omega}\intx\left[e^2\vec a_\bot^2
+\lambda(h^2+\varphi^2)+2\lambda h^2\right]\\
&&-\frac{e^2}{\Omega}\phi\intxy \pi_\varphi(\vec x) G(\vec x-\vec y)\pi_\varphi(\vec y)
\nonumber\\
&&+\frac{e^2}{2\Omega}\Pi\intxy\left[\pi_\varphi(\vec x) G(\vec x-\vec y)
\varphi(\vec y)+\varphi(\vec x) G(\vec x-\vec y)\pi_\varphi(\vec y)\right]\pkt
\nonumber
\eea 
We also need the equations of motion for the three different 
quantum fluctuations. The first one is the transverse gauge field. We find
an equation for the field $a_{\bot i}$ itself and for its canonical momentum.
We can combine these two expressions to a second order differential equation
for the gauge field:
\bea
&&\dot a_{\bot i}=\pi_{\bot i}\kma\hspace{0.5cm}\dot \pi_{\bot i}
=(\triangle-e^2\phi^2)a_{\bot i}\nonumber\\
&&\Rightarrow \ddot a_{\bot i}=(\triangle -e^2\phi^2)a_{\bot i}\pkt
\eea
In the same way, we get the equation of motion for the real component
of the Higgs fluctuation
\be
\ddot h=\left[\triangle-\lambda(3\phi^2-v^2)\right]h\pkt
\ee
More difficulties arise for the Goldstone sector $\varphi$ because the 
field and its canonical momentum are coupled via Green functions. We find 
for the field itself 
\be
\dot \varphi=\pi_\varphi+\frac{e^2}{2} \intx\left[\phi G_{xy}(\phi\pi_\varphi-\Pi
\varphi)+(\phi\pi_\varphi-\Pi\varphi)G_{xy}\phi\right]\kma
\ee
and for the momentum
\be
\dot \pi_\varphi=\triangle\varphi-\lambda\varphi(\phi^2-v^2)
-\frac{e^2}{2} \intx\left[\Pi^2 G_{xy}\varphi-(\Pi\phi+\phi\Pi)
G_{xy}\pi_\varphi\right]\pkt
\ee
Now we  transform the equations of motion for the zero mode and the fluctuations
into Fourier space. 
Therefore, we expand the fluctuation fields in the following way 
\be
\label{expansion}
\psi(\vec x,t)=\intk\frac{1}{2\omega_0}\left\{a_kU(k,t)e^{i\vec k\cdot\vec x
-i\omega_0 t}+a^\dagger_k U^*(k,t)e^{-i\vec k\cdot\vec x+i\omega_0 t}\right\}\kma
\ee
with the usual commutator relations for the annihilation and creation
operators $a_k,a^\dagger_k$:
\be
[a_k,a^\dagger_{k'}]=(2\pi)^3 2\omega_0\delta^3(\vec k-\vec k')\pkt
\ee
The $U(k,t)$ are the mode functions for the fluctuations depending on $k$.
For convenience we omit the momentum dependence in the following.
The Fourier transform for the Green function leads to a factor $1/k^2$.
With these expansions, the equation of motion for the zero mode reads
\bea
\label{phidot}
\dot\phi(t)&=&\Pi(t)\left[1+e^2\intk
\frac {1}{2\omega_{p0}k^2}|U_{\varphi}(t)|^2\right]
\nonumber\\
&&-\frac{e^2}{2}\phi(t)\intk\frac{1}{2\omega_{p0}k^2}
\left[U_{\varphi}(t)U^*_{\pi_\varphi}(t)+U_{\pi_\varphi}(t)U_{\varphi}^*(t)\right]\kma
\eea
and for the canonical momentum
\bea
\label{pidot}
\dot \Pi(t)&=&-U'(\phi)\nonumber\\
&&-\phi(t)\intk\left[\frac{e^2}{\omega_{a0}}|U_\bot(t)|^2
+\frac{3\lambda}{2\omega_{h0}}|U_{h}(t)|^2+
\frac{\lambda}{2\omega_{p0}}|U_{\varphi}(t)|^2+\frac{e^2}{2\omega_{p0}
k^2}|U_{\pi_\varphi}(t)|^2\right]\nonumber\\
&&+\Pi(t)\intk\frac{e^2}{4\omega_{p0}k^2}
\left[U_{\pi_\varphi}(t)U_{\varphi}^*(t)+U_{\pi_\varphi}^*(t)U_{\varphi}(t)\right]\kma
\eea
where we have introduced the following frequencies
\bea
\omega_{p0}^2&=&\left[\vec k^2+\lambda(\phi_0^2-v^2)\right]
(\vec k^2+e^2\phi_0^2)/k^2=\omega_{\varphi 0}^2\omega_{a0}^2/ 
k^2\kma\\
\omega_{\varphi 0}^2&=&\vec k^2+\lambda(\phi_0^2-v^2)\kma\\
\omega_{a0}^2&=&\vec k^2+m_W^2+e^2(\phi_0^2-v^2)\kma\\
\omega_{h0}^2&=&\vec k^2+m_h^2+3\lambda(\phi_0^2-v^2)\kma\\
m_h^2&=&2\lambda v^2\kma~~~m_W^2=e^2 v^2\pkt
\eea
The index 0 indicates the choice of $t=0$. Notice, that we have included
a factor two for the two transverse gauge freedoms in the equation of 
motion for $\dot\Pi$.
$\omega_{p0}$ was introduced in \cite{Boyanovsky:1996dc}.
By comparing these results with the zero mode equation in the 
gauge fixed theory which we have examined in \cite{Baacke:1997kj}
we find some analogies in the fluctuation integrals.
The transverse gauge field and the Higgs field component $h$ lead to 
the same contribution in both theories. The Goldstone channel $\varphi$
fulfills a coupled differential equation. In the gauge invariant description
we have a coupling between the field itself and its canonical momentum. In the 
't Hooft-Feynman-gauge, it couples to the time component of the 
gauge field. This component do not appear in the new description 
because they are unphysical. In the 't Hooft-Feynman gauge the unphysical
degrees of freedom are compensated by the ghosts.

In the same way we have found the zero mode equation, 
we can derive the mode functions for the fluctuation 
fields. We find
\bea
\label{trans}
\left[\frac{d^2}{dt^2}+\omega_a^2(t)\right]U_\bot(t)&=&0\kma\\
\label{isosc}
\left[\frac{d^2}{dt^2}+\omega_{h}^2(t)\right]U_{h}(t)&=&0\kma\\
\label{phi}
\left[\frac{d}{dt}+\frac {e^2}{k^2}\Pi(t)\phi(t)\right] U_{\varphi}(t)
-\frac{\omega_a^2(t)}{k^2} U_{\pi_\varphi}(t)&=&0\kma\\
\label{pi}
\left[\frac{d}{dt}-\frac{e^2}{k^2}\Pi(t)\phi(t)\right]U_{\pi_\varphi}(t)
+\left[\omega_\varphi^2(t)+\frac{e^2}{k^2}\Pi^2(t)\right]
U_{\varphi}(t)&=&0\pkt
\eea
In the mode equations for $U_\varphi$ and $U_{\pi_\varphi}$, a problem
in the IR region becomes obvious. The denominator
with $k^2$  leads to problems for vanishing momentum.
During the discussion of the comparison of the different approaches in section
\ref{equiv}, it will become clear that this IR-instability is caused
by higher loop effects. It is possible by combining the differential equation
for the field (\ref{phi}) and its conjugate momentum (\ref{pi}) to find
an IR-stable mode equation by neglecting all terms of higher order than one loop.

\medskip
The energy density can easily be calculated from the Hamiltonian. We have to
insert the expansion of the fields in dependence of the mode functions with
the corresponding creators and annihilators. Then we have to take the
expectation value in the ground state. The field expansion is analogous to
the one for the equation of motion (\ref{expansion}). We find for the energy
density in the Fourier space
\bea
{\cal E}&=&\frac 1 2\Pi^2(t)+U[\phi(t)]\nonumber\\
&&+\intk\frac{1}{2\omega_{a0}}\left[|\dot U_\bot(t)|^2+\omega_a^2(t)
|U_\bot(t)|^2\right]\nonumber\\
&&+\intk\frac{1}{4\omega_{h0}}\left[|\dot U_{h}(t)|^2
+\omega_{h}^2(t)|U_{h}(t)|^2\right]\nonumber\\
&&+\intk\frac{1}{4\omega_{p0}}\left[|U_{\pi_\varphi}(t)|^2
+\omega_{\varphi }^2(t)|U_{\varphi}(t)|^2\right]\nonumber\\
&&+\intk\frac{e^2}{4\omega_{p0}k^2}\left\{\phi^2(t)
|U_{\pi_\varphi}(t)|^2+\Pi^2(t)|U_{\varphi}(t)|^2\right.\nonumber\\
&&\left.\hspace{2cm}-\Pi(t)\phi(t)\left[U_{\pi_\varphi}^*(t)U_{\varphi}(t)
+U_{\pi_\varphi}(t)U_{\varphi}^*(t)\right]\right\}\pkt
\eea
It is easy to check the conservation of the energy by determining the time
derivative. By using the equations of motion we can show that it vanishes.

\section{Coulomb Gauge}
\subsection{Equation of Motion and Energy Density}
Our starting point is again the Lagrangian (\ref{lagrange}).
As discussed, e.g. in \cite{Boyanovsky:1996dc}, and in the last section
it is possible
to find a gauge invariant description of the Abelian Higgs model
by quantizing the theory with Dirac's method. 
An equivalent way is to choose the Coulomb gauge condition 
$\vec\nabla\cdot\vec A=0$. One gets a Hamiltonian written in terms
of transverse components and including the instantenous 
Coulomb interaction. This Coulomb interaction can be traded
with a Lagrange multiplier field linearly coupled to the charge density.
This leads to the Lagrangian in the Coulomb gauge
\bea
\label{lagr}
\call&=&\frac 1 2 \partial_\mu\Phi^\dagger\partial^\mu\Phi+\frac 1 2
\partial_\mu\vec A_T\partial^\mu\vec A_T-e\vec A_T\cdot\vec j_T
-e^2\vec A_T\cdot\vec A_T\Phi^\dagger\Phi\nonumber\\
&&+\frac 1 2 (\nabla A_0)^2+e^2A_0^2\Phi^\dagger\Phi-ieA_0\rho
-V(\Phi^\dagger\Phi)\kma
\nonumber\\
\vec j_T&=&i(\Phi^\dagger\vec\nabla_T\Phi-\vec\nabla_T\Phi^\dagger\Phi)\kma
~~~\rho=-i(\Phi\dot\Phi^\dagger-\Phi^\dagger\dot\Phi)\kma
\eea
where $A_T$ is the transverse component of the gauge field.
The field $A_0$ is a gauge invariant Lagrange multiplier whose
equation of motion is algebraic:
\be
\nabla^2 A_0(\vec x,t)=\rho(\vec x,t)\pkt
\ee
Using the usual decomposition of $\Phi$ into an expectation value
and a fluctuation part, and splitting the fluctuations into a real part
$h$ and an imaginary part $\varphi$,
we can derive the equations of motions for the different fluctuations
and the zero mode. We give the equations in the Fourier space. 
The physical degrees of freedom - the transversal gauge mode and the 
Higgs mode $h$ - are the same as in the gauge invariant approach and
in the 't Hooft-Feynman gauge.
We now  investigate the remaining degrees
of freedom, the  Goldstone field $\varphi$ and the Lagrange multiplier
$a_0$. As already mentioned,
the equation that fixes $a_0$ is purely algebraic. The field equation
reads
\be
\label{alga}
\omega_a^2(t)a_0(t)=e\left[\dot \varphi(t)\phi(t)-\varphi(t)
\dot \phi(t)\right]\pkt
\ee 
For the Goldstone field we find
\be
\label{uvarphi}
\ddot \varphi(t)+\left[
k^2+\lambda (\phi^2(t)-v^2)\right]\varphi(t)
=e\dot a_0(t)\phi(t)+2ea_0(t)\dot \phi(t)\pkt
\ee
It is possible  to eliminate the mode function for
the Lagrange multiplier $a_0$ in (\ref{uvarphi}). By using the differential
equation for $\varphi$ and the classical equation of motion 
\be
\label{Hem}
\ddot \phi-\lambda\phi(\phi^2-v^2)=0\kma
\ee
it is easy to show that the time derivative of $a_0$ is given by
\be
\label{algad}
\dot a_0(t)=\frac{e}{\omega_a^2(t)}\left[
\phi(t)\ddot \varphi(t)-\varphi(t)\ddot \phi(t)-2e\phi(t)\dot \phi(t)a_0(t)\right]
=-e\phi(t)\varphi(t)\pkt
\ee
Inserting (\ref{alga}) and (\ref{algad}) in (\ref{uvarphi}) leads to
\be
\label{coulch}
\calm_{\varphi\varphi}(t)\varphi(t)=0\kma
\ee
where $\calm_{\varphi\varphi}$ is given by 
\be
\label{mphiphi}
\calm_{\varphi\varphi}(t)=\frac{d^2}{dt^2}+k^2+e^2 \phi^2(t)
+\lambda \left[\phi^2(t)-v^2\right]-\frac{2e^2\dot \phi(t)}{\omega_a^2(t)}
\left[\phi(t)\frac{d}{dt}-\dot \phi(t)\right]\pkt
\ee
$a_0$ is a dependent 
mode and the physical modes are the two transversal gauge fields
$a_\bot$ and the scalar Higgs mode $h$. 
From the Lagrangian (\ref{lagr}), we can also derive the equation of motion 
for the zero mode $\phi$. We find
\begin{eqnarray}
\label{couleq}
\ddot \phi+\lambda \phi(\phi^2-v^2)&+&3\lambda \phi\langle h^2\rangle
+2e^2\phi\langle a_\bot^2\rangle+\lambda \phi\langle\varphi^2\rangle
\nonumber\\
&&-e^2 \phi\langle a_0^2\rangle+2e\langle a_0\dot\varphi\rangle
+e\langle\dot a_0\varphi\rangle=0\kma
\end{eqnarray}
where the expectation values for the fields and their normalization
are not specified yet. Now we  eliminate the field $a_0$
{\em without} using the classical equation of motion. We set $\ddot \phi$
equal to
\begin{equation}
\label{classic}
\ddot \phi=-\lambda \phi(\phi^2-v^2)+\calr(t)\kma
\end{equation}
where $\calr(t)$ contains the fluctuations. The field equations
are then given by
\begin{eqnarray}
\label{anull}
a_0&=&\frac{e}{\omega_a^2}(\phi\dot\varphi-\dot \phi\varphi)\kma\\
\label{anulld}
\dot a_0&=&-e\phi\varphi+\frac{e}{k^2}\varphi {\cal R}\kma\\
\label{phidd}
\ddot\varphi&=&-\omega_{e\lambda}^2\varphi+\frac{2e^2\dot \phi}{\omega_a^2}
(\phi\dot\varphi-\dot \phi\varphi)+\frac{e^2}{k^2}\phi{\cal R}\varphi\kma
\end{eqnarray}
with
\begin{equation}
\omega_{e\lambda}^2=k^2+e^2\phi^2+\lambda(\phi^2-v^2)\pkt
\end{equation}
By taking the fluctuation integral $\calr(t)$ into account, we obtain a mode
function which is of order higher than one loop. In this case we are dealing
with the back reaction of the quantum fluctuations $a_0$ and $\varphi$.
Obviously, these higher loop terms contain a factor $e^2/k^2$ as in the
Hamiltonian approach which lead to the IR-instabilities. By choosing
$\calr(t)=0$ we consider only one loop effects and
 (\ref{phidd}) is identical with (\ref{coulch}). We will discuss this case
in detail at the end of this section.
Using (\ref{anull}) and (\ref{anulld}), we find for the zero
mode the equation of motion in terms of the quantum fluctuations
\begin{eqnarray}
\label{coulcl}
\ddot \phi+\lambda \phi(\phi^2-v^2)&+&3\lambda \phi\langle h^2\rangle
+2e^2\phi\langle a_\bot^2\rangle\nonumber\\
&+&\left(\lambda-e^2\right)\phi\langle\varphi^2\rangle
-{e^4\phi\dot \phi^2}\langle\frac{\varphi^2}{\omega_a^4}\rangle\nonumber\\
&+&e^2\phi\langle\frac{2\omega_a^2-e^2\phi^2}{\omega_a^4}
\dot\varphi^2\rangle
\nonumber\\
&-&2e^2\dot\phi^2
\langle\frac{k^2}{\omega_a^4}\varphi\dot\varphi\rangle
+e^2\calr\langle\frac{\varphi^2}{k^2}\rangle=0\pkt
\end{eqnarray}
Again we see here the appearance of a higher loop contribution which is IR-divergent.
We can also compute the energy density in the Coulomb gauge.
It is given by
\begin{eqnarray}
\label{coulen}
\cale=\frac 1 2 \dot \phi^2+\frac\lambda 4\left(\phi^2-v^2\right)^2
&+&\frac 1 2\left[\langle\dot h^2\rangle
+\langle\omega_h^2 h^2\rangle\right]
+\left[\langle\dot a_\bot^2\rangle+\langle\omega^2_a
a_\bot^2\rangle\right]\nonumber\\
&+&\frac 1 2 \left[\langle\dot\varphi^2\rangle
+\langle\omega^2_\varphi\varphi^2\rangle\right]
-\frac 1 2\langle\omega_a^2 a_0^2\rangle\pkt
\end{eqnarray}
The component $\langle a_0^2\rangle$ has a negative sign because
of the indefinite metric of the time component of the gauge field.
By inserting (\ref{anull}) into the last term of the energy,
$\cale$ can be written similar to the equation of motion in the form
\begin{eqnarray}
\label{energy}
\cale=\frac 1 2 \dot \phi^2+\frac\lambda 4\left(\phi^2-v^2\right)^2
&+&\frac 1 2\left[\langle\dot h^2\rangle
+\langle\omega_h^2 h^2\rangle\right]
+\left[\langle\dot a_\bot^2\rangle+\langle\omega^2_a
a_\bot^2\rangle\right]\nonumber\\
&+&\langle\frac{k^2}{2\omega_a^2}\dot\varphi^2\rangle
+\frac 1 2\langle\left(\omega_\varphi^2
-\frac{e^2\dot \phi^2}{\omega_a^2}\right)
\varphi^2\rangle\nonumber\\
&+&e^2\phi\dot \phi\langle\frac{\varphi}{\omega_a^2}\dot\varphi\rangle
\pkt
\end{eqnarray}
With (\ref{classic}), (\ref{phidd}) and the equations of motion
for $h$ and $a_\bot$, it is straightforward to show that the
time derivative of $\cal E$ vanishes. Now we have to decide
whether we neglect $\calr(t)$ or not. For a well defined one loop
approximation which we have considered in the 't Hooft-Feynman gauge
in \cite{Baacke:1997kj} we have to take it to be zero. Numerically,
this leads to some problems for the energy conservation.
In order to show the energy conservation, we have to compute the time derivative
of $\cale$. In the energy density terms proportional $\dot\phi$ appear,
and, after performing the time derivative, it is necessary to insert
the equation of motion for the zero mode. Since we have taken $\calr(t)=0$,
we have to insert $\ddot\phi=-\lambda\phi(\phi^2-v^2)$. Analytically, this is
no problem, but numerically this equation is only solved at the initial time.
By solving the equation of motion for the zero mode numerically
we automatically
take the fluctuation integral into account. Therefore, we cannot expect
exact energy conservation. On the other hand this is a good cross check for
the quality of the one loop approximation. For small coupling constant $e^2$
the numerical energy conservation has to be acceptable.

\section{Equivalence of the Different Approaches}
\label{equiv}
At this point we can compare the gauge invariant approach
and the Coulomb gauge. 
In the gauge invariant approach the Goldstone channel was
described by a combination of two first order differential equations
for the field itself (\ref{phi}) and its canonical momentum (\ref{pi}). 
Now we  show that this approach also leads in the one loop order
to the same equations as in the Coulomb gauge. If we differentiate
(\ref{phi}) with respect to $t$ and use the classical equation of motion
$\dot \phi=\Pi$, we find a second order differential equation of the form
\begin{equation}
\ddot U_{\varphi}+\frac{e^2}{k^2}\left(\dot\phi^2+\phi\ddot\phi\right)
U_{\varphi}+\frac{e^2}{k^2}\phi\dot\phi\dot U_{\varphi}
-\frac{2e^2\phi\dot\phi}{k^2}U_{\pi_\varphi}-\frac{\omega_a^2}{k^2}\dot U_{\pi_\varphi}=0\pkt
\end{equation}
With the relations for $U_{\pi_\varphi}$ and $\dot U_{\pi_\varphi}$
\begin{eqnarray}
\label{upi}
U_{\pi_\varphi}&=&\frac{k^2}{\omega_a^2}\dot U_{\varphi}+
\frac{e^2\phi\dot\phi}{\omega_a^2}U_{\varphi}\kma\\
\label{upid}
\dot U_{\pi_\varphi}&=&\frac{e^2\phi\dot\phi}{\omega_a^2}\dot U_{\varphi}
+\frac{e^4\phi^2\dot\phi^2}{k^2\omega_a^2}U_{\varphi}
-\omega_\varphi^2U_{\varphi}-\frac{e^2}{k^2}\dot\phi^2U_{\varphi}\kma
\end{eqnarray}
and again with the classical equation of motion now in the form
$\ddot\phi=-\lambda\phi(\phi^2-v^2)$, it is straightforward to show
\begin{equation}
{\cal M}_{\varphi\varphi}U_{\varphi}=0\kma
\end{equation}
where $\calm_{\varphi\varphi}$ is given by (\ref{mphiphi}). 
By inserting the classical equation of motion without the fluctuation part,
we suppress the higher loop terms and therefore get rid of the IR problem.
Therefore,
we have shown that if the classical equation of motion is fulfilled, 
the Goldstone mode in the Coulomb gauge and in
the Hamiltonian approach has the same fluctuation operator.
We also want to compare the equation of motion for the zero mode
$\phi$ in the Coulomb gauge and in the Hamiltonian approach. In order
to shorten the notation and make a comparison easier we also
write here the fluctuation integrals as expectation values
and do not worry about the normalization. By taking the time derivative
of the zero mode equation (\ref{phidot}), we find:
\begin{equation}
\ddot\phi=\dot\Pi+\dot\Pi e^2\langle\frac{\varphi^2}{k^2}\rangle
+2e^2\Pi\langle\frac{\varphi\dot\varphi}{k^2}\rangle
-e^2\dot\phi\langle\frac{\varphi\pi_\varphi}{k^2}\rangle
-e^2\phi\left[\langle\frac{\dot\varphi\pi_\varphi}{k^2}\rangle
+\langle\frac{\varphi\dot\pi_\varphi}{k^2}
\rangle\right]\pkt
\end{equation}
Now we insert the differential equation for the conjugate momentum $\dot\Pi$
(\ref{pidot}). Since we are only interested in the one loop order
we can use the classical equation of motion $\dot\phi=\Pi$ without
fluctuations and neglect terms of higher orders arising from the product
of fluctuation integrals. Then the linearized field equation
reads as a second order differential equation
\begin{eqnarray}
\ddot\phi=-U'(\phi)&-&\phi\left[2e^2\langle a_\bot^2\rangle
+3\lambda\langle h^2\rangle+\lambda\langle\varphi^2\rangle
+{e^2}\langle\frac{\pi_\varphi^2}{k^2}\rangle\right]\nonumber\\
&-&{e^2}U'(\phi)\langle{\varphi^2}{k^2}\rangle+2e^2\dot\phi
\langle\frac{\varphi\dot\varphi}{k^2}\rangle
-e^2\phi\left[
\langle\frac{\dot\varphi\pi_\varphi}{k^2}\rangle
+\langle{\varphi\dot\pi_\varphi}{k^2}\rangle
\right]\pkt
\end{eqnarray}
By using (\ref{upi}) and (\ref{upid}) for the conjugate momentum of the
fluctuation field $\varphi$, we get the same result as in (\ref{coulcl})
if we choose $\calr(t)=0$. 


\section{Renormalization}
We are interested in the behavior of the zero mode under the influence of the quantum
fluctuations and in the energy. Therefore, we need well defined, finite relations,
and we have to renormalize the equation of motion and the energy density.
With a perturbative expansion of the mode functions it is possible to
extract the leading divergences and to introduce counter terms independent
of the initial conditions. This method was developed for non-equilibrium
systems in \cite{Baacke:1997se} and implemented for gauge theories in
\cite{Baacke:1997kj}. We have given some details of the perturbative
expansion of the mode functions in appendix \ref{appA}. In order to make
a comparison with the 't Hooft-Feynman gauge possible, we also introduce
some renormalization conditions. 
We choose as renormalization point the minimum of the effective potential, which is due to
Nielsen's theorem the same point for both gauges.
The renormalization has
to satisfy the condition $|\Phi|=v$ or in other words the one loop
effective potential has to satisfy some basic properties of the classical
potential. These include the same minimum and the same curvature at the minimum
for both potentials. Therefore, we have to add some finite corrections to the Lagrangian.
In general the effective potential is not very useful in the context of 
non-equilibrium dynamics as it is discussed for example in \cite{Boyanovsky:1993vi}.
Especially, the imaginary part of the effective potential leads to instabilities
and to misleading results for the investigations of the dynamics of the system.
Nevertheless, the minimum of the effective potential point seems to be a natural
point for the renormalization since we are using for the non-equilibrium
calculations the same approximation scheme. For our numerical 
investigations we consider only initial conditions near the minimum
of the potential since this is the only stable point for the one loop
approximation. If we would like to enter in the instable region of the potential
then we have to take the back reaction also in the mode functions into account.
This is beyond the scope of this paper. We are  here interested in the
comparison of different gauges and the influence of additional gauge modes in
general. Therefore, it is useful to restrict ourselves to the stable
region of the potential. 

\subsection{Equation of Motion}
First of all, we have to write the equation of motion (\ref{coulcl}) in terms of
the mode functions. Details for the quantization of the Goldstone mode
$\varphi$ are given in appendix \ref{appA}. In order to deal with a 
well defined one loop approximation we choose $\calr$ equals to zero.
The equation of motion for the zero mode then reads
\be
\ddot\phi(t)+\lambda\left[\phi^2(t)-v^2\right]\phi(t)+\calf(t)=0\kma
\ee
with
\bea
\calf(t)&=&3\lambda\phi\intk\frac{|U_h|^2}{2\omega_{h0}}
+2e^2\phi\intk\frac{|U_\bot|^2}{2\omega_{a0}}\\
&&+(\lambda-e^2)\phi\intk\frac{\omega_a^2}{2k^2\tilde\omega_{e\lambda 0}}
|\tilde U_\varphi|^2\nonumber\\
&&+e^2\phi\intk\frac{2k^2+e^2\phi^2}{2k^2\tilde\omega_{e\lambda 0}\omega_a^2}
|\dot{\tilde U}_\varphi|^2
-e^4 \phi\dot\phi^2\intk
\frac{2\omega_a^2+k^2}{2\omega_a^6\tilde\omega_{e\lambda 0}}
|\tilde U_\varphi|^2\nonumber\\
&&+e^2\dot\phi\intk\frac{e^2\phi^2\omega_a^2-k^4}
{2k^2\tilde\omega_{e\lambda 0}\omega_a^4}
\left(\tilde U_\varphi\dot{\tilde U}^*_\varphi
+\tilde U_\varphi^*\dot{\tilde U}_\varphi\right)\nonumber\pkt
\eea
The first four integrals contain divergent contributions, which we can
extract by introducing the perturbative expansion for the mode functions 
as explained in appendix \ref{appA}. Dimensional regularization of the divergent
parts leads to the following counter terms 
\bea
\label{ctm}
\delta m&=&{3\lambda m_h^2}
I_{-1}(m_h)
-2e^2 m_h^2
I_{-3}(m_h)
+(\lambda +e^2)m_h^2
I_{-1}(m_h)
\kma\\
\delta\lambda&=&\left[(\lambda-e^2)^2+9\lambda^2\right]
I_{-1}(m_h)
+e^4I_{-3}(m_h)\kma\\
\label{ctz}
\delta Z&=&2e^2 I_{-3}(m_h)\kma
\eea
with
\bea
I_{-1}(m_h)&=&\frac{1}{16\pi^2}
\left\{\frac 2 \epsilon +\ln\frac{4\pi\mu^2}{m_h^2}-\gamma+1\right\}\kma\\
I_{-3}(m_h)&=&\frac{1}{16\pi^2}
\left\{\frac 2 \epsilon +\ln\frac{4\pi\mu^2}{m_h^2}-\gamma\right\}\pkt
\eea
After introducing the counter terms we finally get the following finite
equation of motion
\bea
(1+\delta Z)\ddot \phi-\frac 1 2 (m_h^2+\Delta m^2)\phi+(\lambda+\Delta \lambda)\phi^3
+\calf_{\rm fin}=0\kma
\eea
with finite corrections for the wave function, the mass, and the coupling constant
\bea
\Delta Z&=&-2e^2 C\kma\\
\Delta m^2&=&\frac{1}{64\pi^2}\left[6\lambda m_h^2+6e^2 m_W^2-2 e^2 m_h^2
+m_h^2(e^2-\lambda)\ln\frac{2\lambda}{e^2}\right.\nonumber\\
&&\hspace{1.5cm}+3\lambda m_h^2\ln\frac{m_h^2}{m_{h0}^2}
+(\lambda-e^2)m_h^2\ln\frac{m_h^2}{m_{\varphi 0}^2}\\
&&\hspace{1.5cm}-4 e^2\phi_0^2
-18\lambda^2\phi_0^2-2(\lambda+e^2)^2\phi_0^2\biggr]\kma\\
\Delta\lambda&=&\frac{1}{64\pi^2}\left[4\lambda(\lambda+e^2)
+(e^2-\lambda)^2\ln\frac{2\lambda}{e^2}\right.\nonumber\\
&&\hspace{1.5cm}-9\lambda^2\ln\frac{m_h^2}{m_{h0}^2}-2e^4\ln\frac{m_W^2}{m_{W0}^2}
\left.-(\lambda-e^2)^2\ln\frac{m_h^2}{m_{\varphi 0}^2}\right]\nonumber\kma\\
C&=&\frac{1}{16\pi^2}\ln\frac{m_h^2}{m_{\varphi 0}^2}\kma
\eea
The terms independent of the initial conditions are due to the choice of the
renormalization conditions The finite fluctuation integral reads
\bea
{\cal F}_{\rm fin}&=&
\label{ffin}
3\lambda\phi\intk\frac{1}{2\omega_{h0}}\left(2\re f_h+|f_h|^2\right)
\nonumber\\
&&+2e^2\phi\intk\frac{1}{2\omega_{a0}}\left(2\re f_\bot+|f_\bot|^2\right)
\nonumber\\
&&+e^2\phi\intk\frac{m_W^2\left(m_W^2-\tilde m_{e\lambda 0}^2\right)}
{2k^2\omega_a^2\tilde\omega_{e\lambda 0}}
-e^2\phi\intk\frac{V_{e\lambda}\tilde m_{e\lambda 0}^2}{2k^2\tilde\omega_{e\lambda 0}^3}
\nonumber\\
&&-e^2\phi\intk\frac{m_{W0}^2 m_{\varphi0}^2}{k^2\omega_{a0}^2\tilde\omega_{e\lambda 0}}
\nonumber\\
&&+(\lambda-e^2)\phi\intk\frac{1}{2\tilde\omega_{e\lambda 0}}
\left[\frac{\omega^2_a}{k^2}\left(
2\re f_\varphi+|f_\varphi|^2\right)+
\frac{V_{e\lambda}}{2\tilde\omega^2_{e\lambda  0}}\right]
\nonumber\\
&&+e^2\phi\intk\frac{1}{2k^2\omega_a^2\tilde\omega_{e\lambda 0}}
\left\{\left(2k^2+e^2\phi^2\right)
\left[|\dot f_\varphi|^2
\right.\right.\nonumber\\
&&\hspace{5cm}\left.\left.-\tilde\omega_{e\lambda 0}^2\left(
2\re f_\varphi+|f_\varphi|^2\right)\right]
-\omega_a^2 V_{e\lambda}\right\}
\nonumber\\
&&-e^4 \phi\dot\phi^2\intk
\frac{2\omega_a^2+k^2}{2\omega_a^6\tilde\omega_{e\lambda 0}}
|\tilde U_\varphi|^2\nonumber\\
&&+e^2\dot\phi^2\intk\frac{e^2\phi^2\omega_a^2-k^4}
{2k^2\tilde\omega_{e\lambda 0}\omega_a^4}
\left(\tilde U_\varphi\dot{\tilde U}^*_\varphi
+\tilde U_\varphi^*\dot{\tilde U}_\varphi\right)\pkt
\eea
Therefore, we have a well defined finite equation of motion which we can investigate numerically.

\subsection{Energy Density}
In order to find the divergent parts of the energy density we have to introduce 
the mode functions for the different fields in Eq. (\ref{energy}) in the same way as for the equation
of motion. This leads to
\bea
\cale&=&\label{enmod}
\frac 1 2\dot\phi^2+\frac \lambda 4\left(\phi^2-v^2\right)^2
+\intk\frac{1}{4\omega_{h0}}\left[|U_h^2|^2+\omega_h^2(t)|U_h^2|^2\right]
\nonumber\\
&&+\intk\frac{1}{2\omega_{a0}}\left[|U_\bot^2|^2+\omega_a^2(t)|U_\bot^2|^2\right]\nonumber\\
&&+\intk\frac{\omega_a^2}{4k^2\tilde\omega_{e\lambda 0}}\left(
\omega_\varphi^2-\frac{e^2\dot\phi^2}{\omega_a^2}
\right)|\tilde U_\varphi|^2+\intk\frac{|\dot{\tilde U}_\varphi|^2}{4\tilde\omel}
\\
&&+e^4\phi^2\dot\phi^2\intk\frac{2\omega_a^2+k^2}{4\tilde\omega_{e\lambda 0} k^2\omega_a^4}
|\tilde U_\varphi|^2
\nonumber\\
&&+e^2\phi\dot\phi\intk\frac{2k^2+e^2\phi^2}{4k^2\tilde\omega_{e\lambda 0} \omega_a^2}
\left(\tilde U_\varphi
\dot{\tilde U}_\varphi^*+\tilde U_\varphi^*\dot{\tilde U}_\varphi
\right)\nonumber\pkt
\eea
As in the equation of motion the first four integrals contain
divergences. With the help of the truncated mode functions $f_h$, $f_\bot$, and
$f_\varphi$ we find after a lengthy but straightforward calculation in addition
to the mass and coupling constant counter term a cosmological constant counter
term which arise from a quartic divergence of the form
\be
\label{cosm}
\delta\Lambda=\frac{m_h^4}{128\pi^2}\left\{\frac 2 \epsilon+\ln\frac{4\pi\mu^2}{m_h^2}
-\gamma+\frac 3 2 \right\}\pkt
\ee
We also get an additional finite term 
\bea
\Delta \Lambda&=&-\frac {1}{256\pi^2}\left(5 m_h^4+6 m_W^4-4m_h^2 m_W^2
-m_h^4\ln\frac{2\lambda}{e^2}\right)\nonumber\\
&&-\frac{m_h^4}{128\pi^2}\ln\frac{m_h^2}{m_{h0}^2}\\
&&+\frac{1}{128\pi^2}\left\{\left[2e^4+9\lambda^2+(e^2+\lambda)^2\right]\phi_0^4
+m_h^2(e^2+4\lambda)\phi_0^2\right\}\nonumber
\kma
\eea
so that the complete finite energy density is given by
\bea
\cale_{\rm fin}&=&\frac 1 2 \left(1+\Delta Z\right)\dot\phi^2
-\frac 1 4 \left(m_h^2+\Delta m^2\right)\phi^2
+\frac 1 4\left(\lambda+\Delta\lambda\right)\phi^4+\Delta\Lambda+\Delta\Lambda'
\nonumber\\
&&+\intk\frac{1}{4\omega_{h0}}\left[|\dot f_h|^2+V_h\left(2\re f_h+|f_h|^2\right)
+\frac{V_h^2}{16\omega_{h0}^3}\right]\nonumber\\
&&+\intk\frac{1}{2\omega_{a0}}\left[|\dot f_\bot|^2
+V_a\left(2\re f_\bot+|f_\bot|^2\right)
+\frac{V_h^2}{16\omega_{h0}^3}\right]\nonumber\\
&&+\intk\frac{1}{4\tilde\omega_{e\lambda 0}}\left(
\frac{m_W^2(t)m_\varphi^2(t)}{k^2}+\frac{m_{W0}^2m_{\varphi 0}^2}
{\omega_{a0}^2}\right)\left(
2\re f_\varphi+|f_\varphi|^2\right)
\nonumber\\
&&+\intk\frac{1}{4\tilde\omega_{e\lambda 0}}\left(|\dot f_\varphi|^2
-\frac{V_{e\lambda}^2}{4\tilde\omega_{e\lambda 0}^2}\right)\nonumber\\
&&+\intk\frac{V_{e\lambda}}{4\tilde\omega_{e\lambda 0}}
\left(2\re f_\varphi+|f_\varphi|^2+\frac{V_{e\lambda}}{2\tilde\omega_{e\lambda 0}^2}
\right)\nonumber\\
&&-e^2\dot\phi^2\intk\frac{1}{4\tilde\omega_{e\lambda 0} k^2}
\left(
2\re f_\varphi+|f_\varphi|^2\right)
\nonumber\\
&&+e^4\phi^2\dot\phi^2\intk\frac{2\omega_a^2+k^2}{4\tilde\omega_{e\lambda 0} k^2\omega_a^4}
|\tilde U_\varphi|^2\nonumber\\
&&+e^2\phi\dot\phi\intk\frac{2k^2+e^2\phi^2}{4k^2\tilde\omega_{e\lambda 0} \omega_a^2}
\left(\tilde U_\varphi
\dot{\tilde U}_\varphi^*+\tilde U_\varphi^*\dot{\tilde U}_\varphi
\right)\kma
\eea
which can now be treated numerically. 


\section{Numerical Results}
In order to investigate the influence of the gauge field sector on the
zero mode in a system out of equilibrium, we have carried out some
numerical examples. We are interested in two different aspects: first 
we  investigate the influence of the different gauges. Therefore, we
reinvestigate the 't Hooft-Feynman background gauge which we have
discussed in detail in \cite{Baacke:1997kj}
and we compare the results with those we find in the Coulomb gauge.
Secondly, we  investigate the effect of the 
different degrees of freedom which arise in gauge theories.
For the 't Hooft-Feynman gauge we summarize
the results for the equation of motion, the mode functions, and the energy density
which we have published in \cite{Baacke:1997kj}. For more details, especially
in view of the renormalization, the reader is referred to our paper \cite{Baacke:1997kj}.

\subsection{The 't Hooft-Feynman Gauge}
We give a short review on the main steps to derive the equation
of motion and the energy in the 't Hooft-Feynman background gauge.
In opposite to our previous work \cite{Baacke:1997kj} we restrict
ourselves here to the abelian Higgs model. This leads to some
modifications in the degeneracy factors which we have introduced
in \cite{Baacke:1997kj}. The Lagrangian which describes the abelian
Higgs model in the 't Hooft Feynman gauge contains three parts
\be\label{ltot}
\call_{\rm tot}=\call+\call_{\rm GF}+\call_{\rm FP}\pkt
\ee
$\call$ is given again by (\ref{lagrange}). We split the Higgs field
into a mean value and fluctuations
\be
\Phi(\vec x,t)=\left[\phi(t)+h(\vec x,t)+i\varphi(\vec x,t)\right]\kma
\ee
and consider for the gauge field only fluctuations
\be
A_\mu(\vec x,t)=a_\mu(\vec x,t)\pkt
\ee
The gauge fixing term has the following form
\be
\call_{\rm GF}=-\frac 1 2 F^2\kma
\ee
with 
\be
F(a_\mu,\varphi)=\partial_\mu a^\mu+e(\phi+h)\varphi\pkt
\ee
The corresponding Faddeev-Popov Lagrangian which is relevant
for our calculations is
\be
\call_{\rm FP}=\partial_\mu\eta^\dagger\partial \eta
-e^2\phi^2\eta^\dagger\eta
\pkt
\ee
After inserting the expansions for the fields in the total Lagrangian
(\ref{ltot}) and neglecting all terms higher than second order in
the fluctuations we derive the following
equation of motion for the zero mode
\begin{eqnarray}
\ddot\phi+\lambda\phi(\phi^2-v^2)&+&3\lambda\phi\langle h^2\rangle
+e^2\phi\langle a_\bot^2\rangle\nonumber\\
&+&\left(\lambda+e^2\right)
\phi\langle \varphi^2\rangle-e^2\phi\langle a_0^2\rangle
-e\partial_t\langle a_0\varphi\rangle=0\pkt
\end{eqnarray}
In this notation we  have not taken care of the normalization and the renormalization
or the different solutions for the mode functions of the coupled channel.
We only want to give an overview of the equations and the connections
of the fields and do not go into technical details.
As already mentioned, the mode functions for the isoscalar Higgs field and the
transversal gauge field are the same as in the Coulomb gauge (\ref{trans}),
(\ref{isosc}). For the coupled sector $a_0\varphi$, they are given by
\begin{eqnarray}
\left\{\begin{array}{cc}-\partial_t^2-\omega_a^2(t)&g\dot\phi(t)\\
g\dot\phi(t)&\partial_t^2+\omega_{e\lambda}(t)^2\end{array}\right\}
\left\{\begin{array}{c}a_0(t)\\
\varphi(t)\end{array}\right\}=0\pkt
\end{eqnarray}
The operator has two interesting features which distinguishes it from
the single modes, the indefinite metric of the time component of the
gauge field, and the time derivative of the zero mode in
the off diagonal elements which connect the two fields. We find analogous
properties in the Coulomb gauge; the fluctuation part for $a_0$
contributes with a negative sign to the fluctuation integral
(\ref{couleq}), and we find time derivatives of the zero mode
 in the equation of motion for the zero mode (\ref{coulcl})
as well as in the mode function for $\varphi$ (\ref{coulch}). 

In the Feynman gauge, the mode functions for the transversal gauge field,
for the longitudinal gauge field, and for the ghost fields are the same.
Two of the three gauge components are canceled by the ghost fields
and only one degree of freedom is left in contrast to the Coulomb gauge
where we have two transverse gauge components. 
The energy density can be derived by integration of the 
equation of motion for the zero mode or from the corresponding
Hamiltonian of the system. It reads (see also \cite{Baacke:1997kj})
\begin{eqnarray}
\cale&=&\frac 1 2 \dot \phi^2+\frac\lambda 4\left(\phi^2-v^2\right)^2
\\
&&+\frac 1 2\left[\langle\dot h^2\rangle
+\langle\omega_h^2 h^2\rangle\right]
+\frac 1 2\left[\langle\dot a_\bot^2\rangle+\langle\omega^2_a
a_\bot^2\rangle\right]\nonumber\\
&&+\frac 1 2 \left[\langle\dot\varphi^2\rangle
+\langle\omega^2_{e\lambda}\varphi^2\rangle\right]
-\frac 1 2\left[\langle \dot a_0^2\rangle+\langle\omega_a^2 a_0^2\rangle\right]
\nonumber\pkt
\end{eqnarray}
This expression looks very similar to the Coulomb energy (\ref{coulen}) despite the fact
that $a_0$ is dynamical and contributes with a derivate part. 

\subsection{Results}
For our numerical calculations, we have chosen four different sets of
parameters listed in Table \ref{tab4}. The initial value for the zero mode
$\phi$ and the Higgs mass $m_h$ are the same for all sets. 
They are chosen in such a way that the zero mode evolves in the right
minimum of the potential. 
With this choice of initial conditions, the zero mode can not evolve
into the complex part of the effective potential around zero. In this region, 
instabilities increase dramatically and the one loop approximation breaks down
as explained for the $\phi^4$~theory for example in \cite{Boyanovsky:1993vi,Baacke:2000fw}.

\begin{figure}[tb]

\vspace{-2.3cm}

\begin{picture}(13,10)
\put(0,0){\epsfxsize=13cm\epsfbox{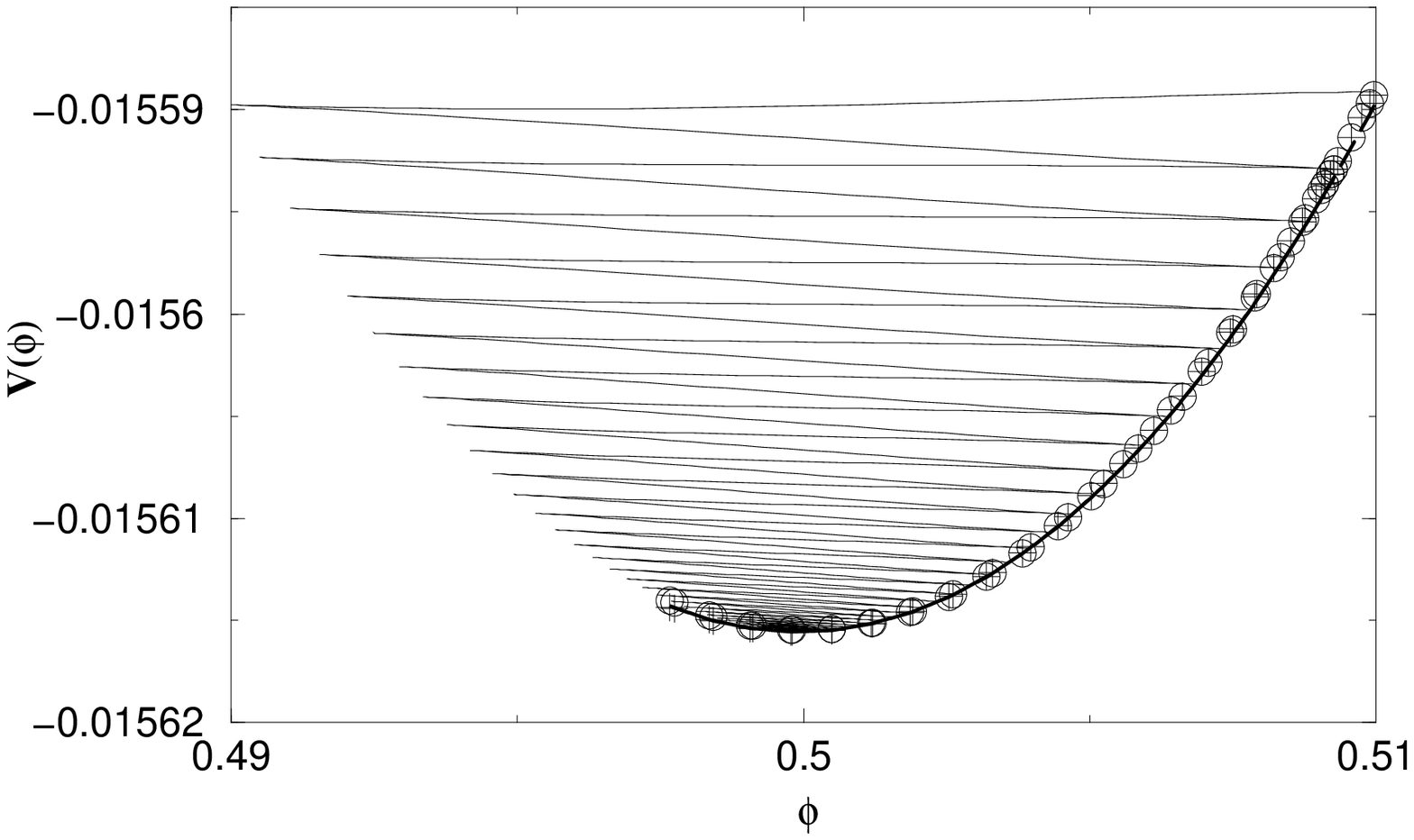}}
\end{picture}
\caption{Potential vs. $\phi$ in the Coulomb gauge for parameter set 4, solid line:
classical energy vs. $\phi$, dashed line: classical potential, circles: classical potential
with finite corrections, pluses: effective potential.}
\label{coulpot}

\vspace{-2cm}

\begin{picture}(13,10)
\put(0,0){\epsfxsize=13cm\epsfbox{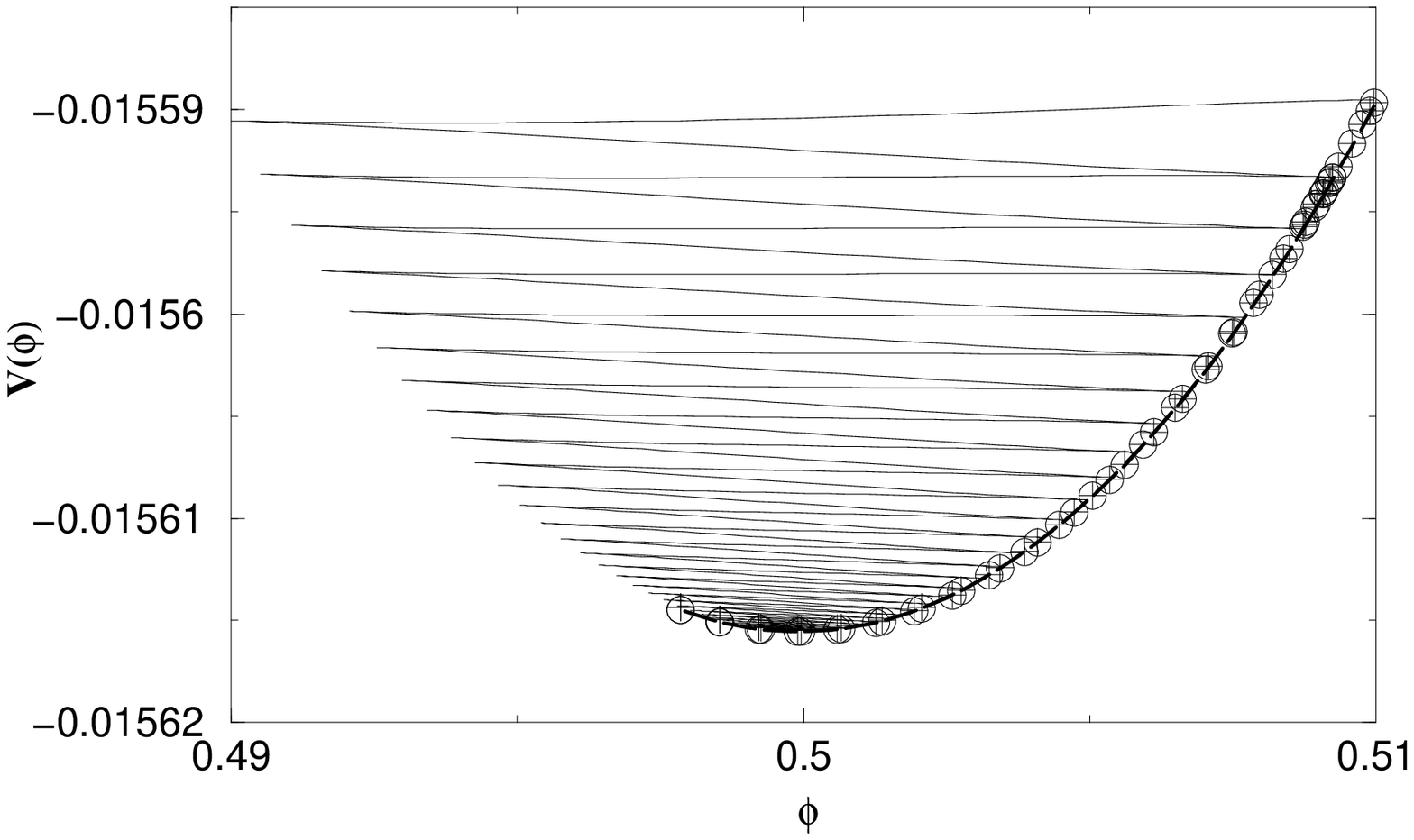}}
\end{picture}
\caption{Potential vs. $\phi$ in the Feynman gauge for parameter set 4, solid line:
classical energy vs. $\phi$, dashed line: classical potential, circles: classical potential
with finite corrections, pluses: effective potential.}
\label{feynpot}
\end{figure}

In order to give an impression of the potential
we have plotted  in Fig.~\ref{coulpot} for the Coulomb gauge 
and Fig.~\ref{feynpot} for the Feynman gauge 
different approximations for the 
potential. The dashed line shows the classical potential
$V(\phi)=\frac{\lambda}{4}(\phi^2-v^2)^2$, the circles include finite corrections
due to the renormalization, and the pluses display the effective one loop potential.
Since we have chosen the renormalization conditions in such a way, that
the classical and the effective potential have the same minimum
and the same curvature at the minimum, they are very close to each other.
The solid line shows the zero mode part of the energy versus $\phi$.
The field begins to roll down the potential but the energy
is not high enough for the field to reach the maximum at zero. Therefore,
it starts to oscillate in the minimum. 
The prediction of these two plots
in the context of non-equilibrium dynamics is not clear, 
because the effective potential is an equilibrium quantity, since the expectation
value of the scalar field, that serves as order parameter, is space time independent.
Nevertheless, they are instructive to get an idea of the potential by which
the zero mode is influenced.

For our numerical considerations  we have only varied
the coupling constants $\lambda$ and $e$  and therefore the masses
of the different fields. We have also given the initial masses 
of the three different fields in Table \ref{tab4}. 

\begin{figure}[tb]

\vspace{-2.3cm}

\begin{picture}(13,10)
\put(0,0){\epsfxsize=13cm\epsfbox{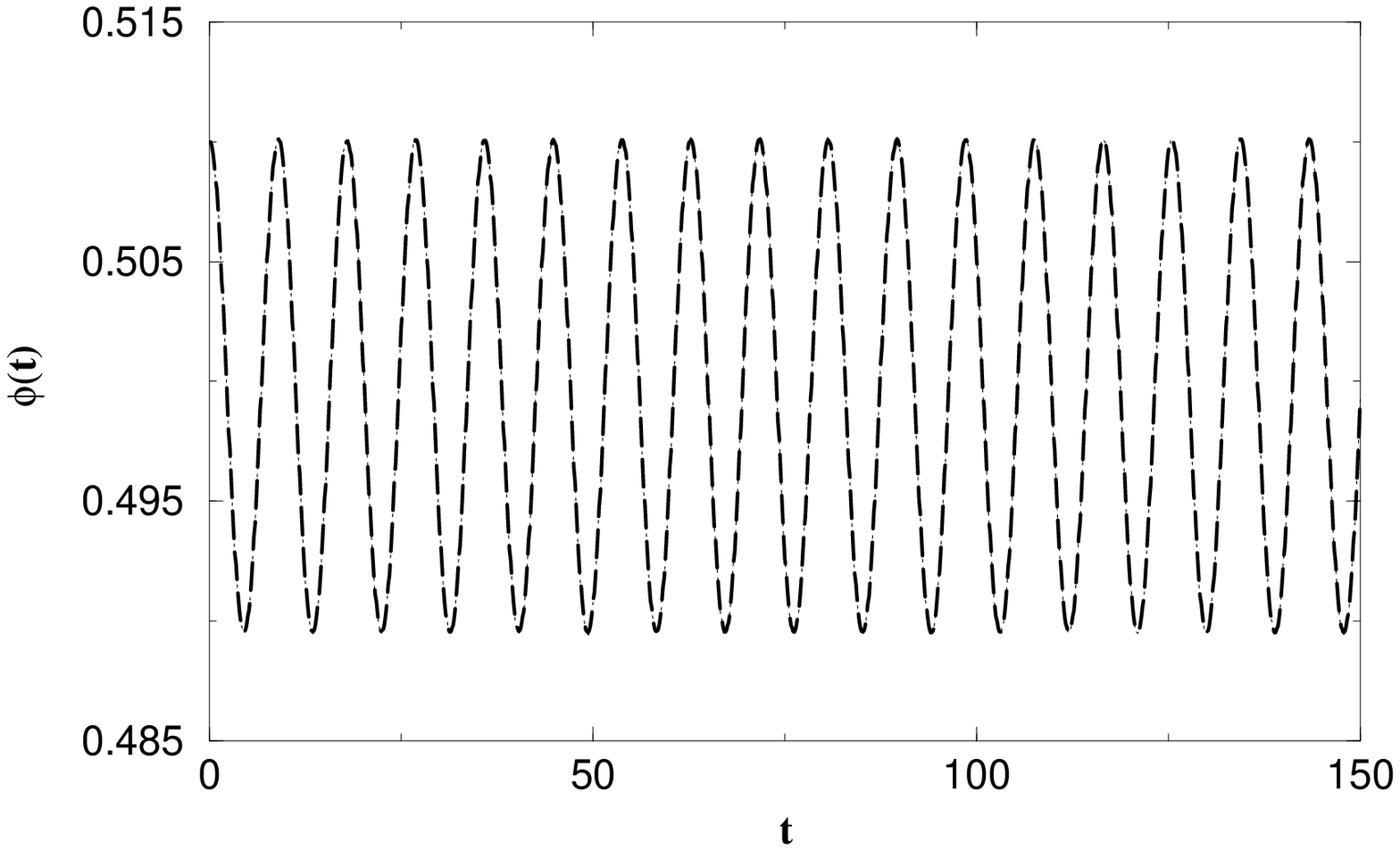}}
\end{picture}
\caption{Zero mode vs. $t$ for parameter set 1, dashed line: Feynman gauge, dotted line: Coulomb gauge.}
\label{l1e1}

\vspace{-1.5cm}

\begin{picture}(12,10)
\put(0.5,0){\epsfxsize=12.5cm\epsfbox{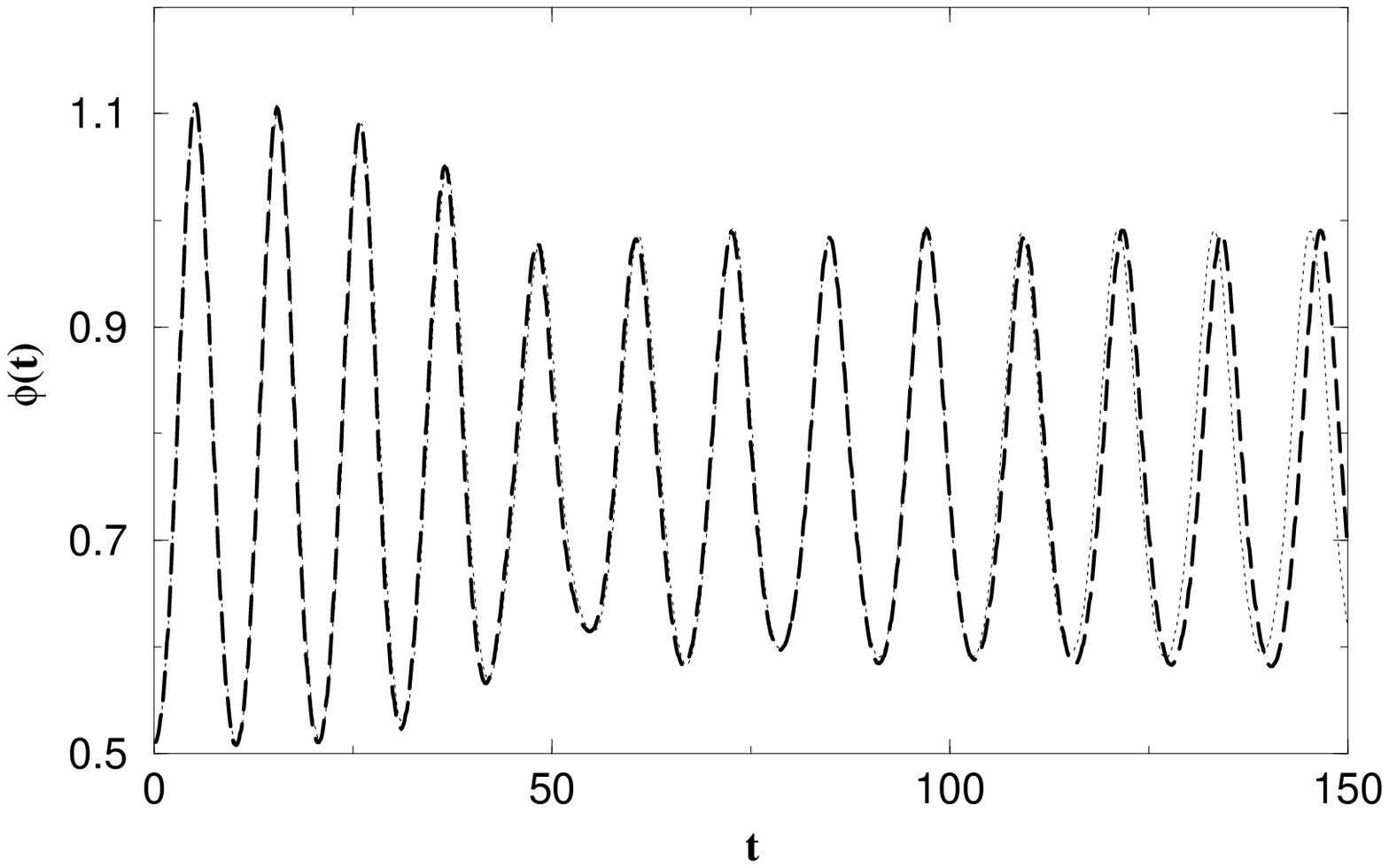}}
\end{picture}
\caption{Zero mode vs. $t$ for parameter set 2, dashed line: Feynman gauge, dotted line: Coulomb gauge.}
\label{l33e13}
\end{figure}

For the first parameter set, we have
chosen the same coupling constant for the Higgs field and the gauge field. The
initial masses for the fields are all small but not zero. Since we have
taken the initial value for the zero mode to be small, the effect of the
quantum fluctuations is negligible and therefore, $\phi$ oscillates with constant
amplitude. The behavior of the zero mode, which is displayed
in Fig.~\ref{l1e1} agrees in both gauges excellent.

In the second parameter set we have chosen a dominant gauge coupling,
therefore the initial mass of the Higgs field becomes very small compared
to $m_{a0}^2$ and $m_{e\lambda 0}^2$. In this case, the zero mode is moderately damped,
as shown in Fig.~\ref{l33e13}; the behavior is comparable to a purely scalar theory. The two gauges
agree again very well, at late time the zero modes are slightly dephased.

The situation changes drastically, if we choose a smaller gauge coupling
and therefore a nearly vanishing Goldstone mass. In Fig.~\ref{inst}
and Fig.~\ref{l1e01} we show two examples for this constellation.

In the first example we have chosen a nearly vanishing gauge coupling
with $e=0.001$. In this case an instability occurs and the system breaks down.
We have analyzed a similar problem occurring in the $\phi^4$ theory in
\cite{Baacke:2000fw}. There we have found that the spontaneous symmetry
breaking leads to instabilities if the mass squared term of the fluctuation
field becomes negative due to the decrease of the zero mode. In addition
to $m_h^2(t)=m_h^2+3\lambda(\phi^2-v^2)$, the mass term, which is also relevant
for the spontaneously broken $\phi^4$~theory, we now have to secure, that
$m_{e\lambda}^2(t)=e^2\phi^2+\lambda(\phi^2-v^2)$ does not become negative when $\phi(t)$
decreases. This leads to the two conditions
\bea
m_h^2(t)>0&\Rightarrow & \phi(t)>\sqrt{\frac{m_h^2}{6\lambda}}\kma\\
m_{e\lambda}^2(t)>0&\Rightarrow & \phi(t)>\sqrt{\frac{m_h^2}{2(e^2+\lambda)}}\pkt
\eea
Therefore, the critical value for the field $\phi(t)$ for parameter set 3 is
around $\phi(t)>0.5$ (exactly 0.49999975) in order to find a stable development
for the system.  We have displayed in the frame the development
of  the zero mode for a shorter time period. The solid line at $0.5$
shows the critical value for $\phi$, for which $m_{e\lambda}^2(t)$ becomes negative. 
One sees that the average value
of the zero mode is slightly smaller than this critical value. Therefore,
the system breaks down.

In the second example we have chosen the gauge coupling to be $e=0.1$,
therefore slightly larger but again small. In this case the zero mode is strongly
damped and settles down to the minimum. 
An appropriate size for the gauge coupling prevents the system to destabilize. 
We can influence $m_{e\lambda}(t)$ either by changing the Higgs coupling
or the gauge coupling. As this example in Fig.~\ref{l1e01}
shows it is possible to find a proper relation between $\lambda$
and $e$ in order to allow the decay of $h$ into $\varphi$ and gauge field and
a stable development of the system in the minimum of the potential.
We have also plotted the critical value for the zero mode
in this case where it is obviously much smaller than the final value
of the zero mode. 

In order to understand this strong damping of the field more precisely,
it is instructive to investigate the Feynman graphs for the problem in detail.
This is especially easy in the 't Hooft-Feynman gauge. We have listed the the occurring
graphs in appendix~\ref{appB}. The relevant graph which leads to the strong damping
is the following

\parbox{15cm}{
\parbox[t]{7cm}
{\begin{center}
\mbox{\epsfxsize=5cm\epsfbox{gekk.eps}}
\end{center}}}

It allows the decay of the Higgs field into $a$ and $\varphi$ for a proper relation between
the masses. In the Coulomb gauge this mechanism is not as obvious as in the 't Hooft-Feynman
gauge since we have eliminated the $a_0$-mode. Also the structure of the fluctuation
integral for the $\varphi$-mode is rather complicated and allows therefore not such a simple
analysis. 

\begin{figure}[tb]

\vspace{-2.3cm}

\begin{picture}(13,10)
\put(0,0){\epsfxsize=13cm\epsfbox{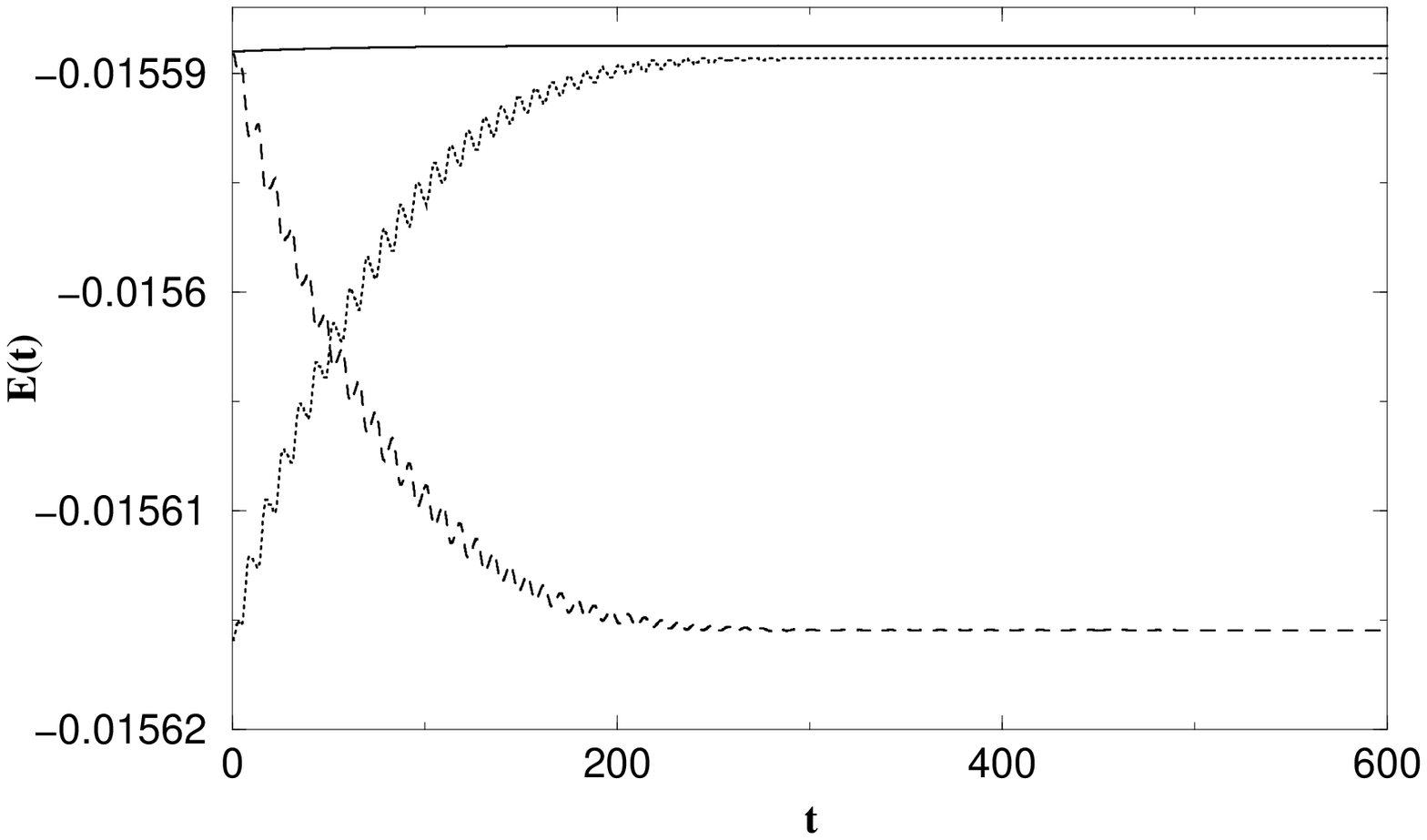}}
\end{picture}
\caption{Energy vs. $t$ in the Coulomb gauge
for parameter set 4, solid line: total energy, 
dashed line: classical energy, dotted line: fluctuation energy.}
\label{coulener}

\vspace{-1.5cm}

\begin{picture}(13,10)
\put(0,0){\epsfxsize=13cm\epsfbox{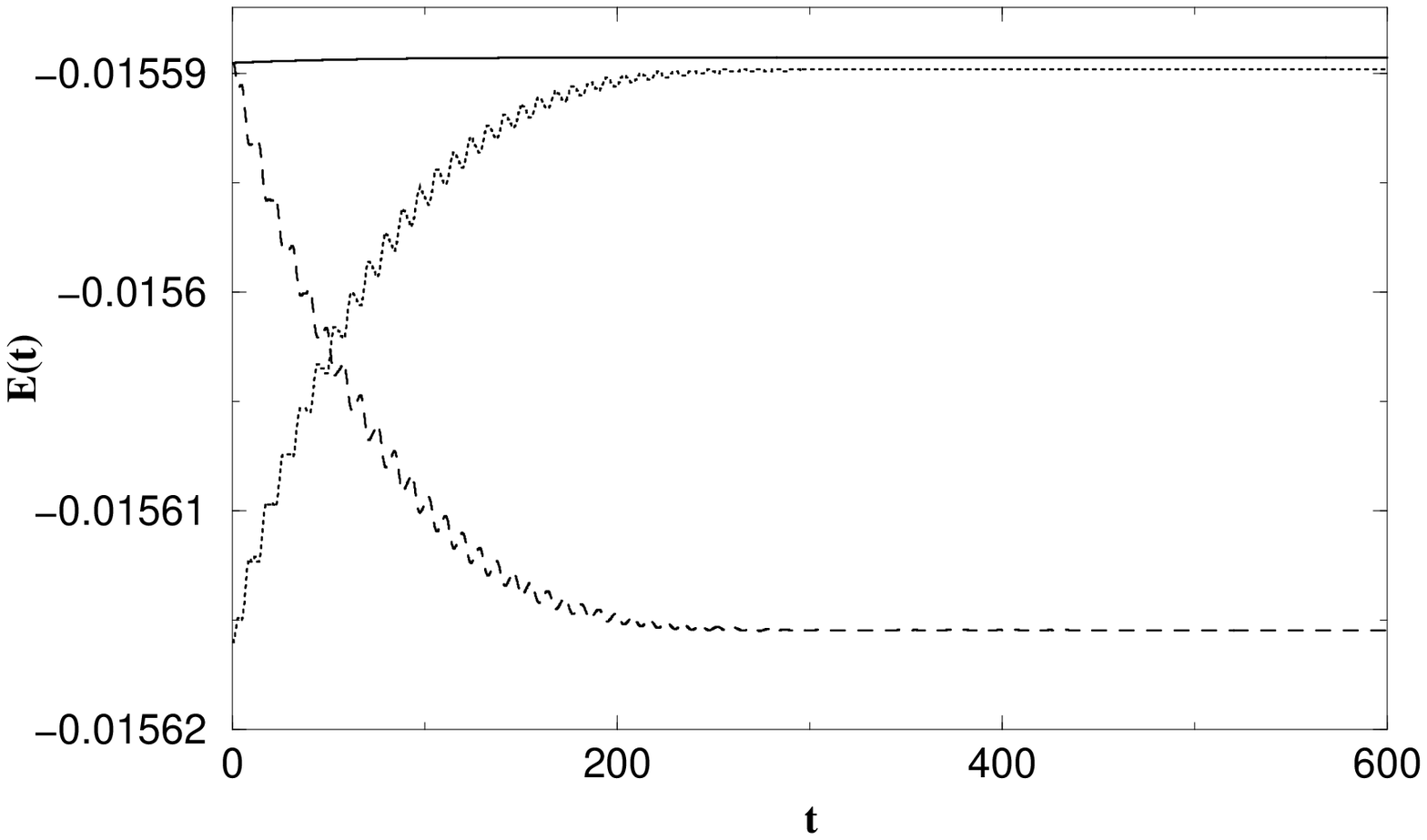}}
\end{picture}
\caption{Energy vs. $t$ in the Feynman gauge
for parameter set 4, solid line: total energy, 
dashed line: classical energy, dotted line: fluctuation energy.}
\label{feynen}
\end{figure}

In order to check our numerics we have plotted the energy density for the Coulomb gauge
in Fig.~\ref{coulener} and the 't Hooft Feynman gauge 
in Fig.~\ref{feynen} for the fourth parameter set.
The upper
line displays the fluctuation energy which increases and the lower line
the zero mode part of the energy which decreases. The solid line shows the total
energy. For convenience we have added in both cases a constant to the
fluctuation part of the energy. Otherwise the curves are only straight 
lines due to their distance.
The energy conservation is excellent in both cases.

Summarizing the results, we have found
that the damping effect is strongest for a nearly massless Goldstone field
and a massive isoscalar Higgs field. 
In this case the isoscalar Higgs field has the possibility to decay
into the other fields. We have found an analogous behavior in the $\phi^4$~theory 
in the large $N$ limit. There, the damping of zero mode
was due to the massless Goldstone bosons. 

We have also found that the behavior of the zero mode and the energy
is the same for the different gauges 
and that the Goldstone channel plays an important role.

\section{Conclusions and Outlook}
We have analyzed in this paper the abelian Higgs model out of equilibrium in detail.
Therefore, we have investigated a gauge invariant approach which was developed
by Boyanovsky et al. \cite{Boyanovsky:1996dc}. We have extended their calculations
to a complete set of equations, which describes the evolution of the zero mode under
the influence of gauge and Higgs fluctuations. We found some problems in the IR region
induced by the inclusion of terms higher then one loop order. Since these terms were
included in an uncontrollable manner, we have computed a linearized form of the equations. 
They are equivalent to the Coulomb gauge fixed theory in the one loop approximation.
We have performed the renormalization of the system in order to get finite well
defined equations which we have investigated numerically. We were also interested
in the behavior of the system under the influence of different gauges. Based
on \cite{Baacke:1997kj} we have reinvestigated the 't Hooft-Feynman gauge.
The main differences between the two gauges is that we have eliminated in the Coulomb gauge
all unphysical degrees of freedom while in the 't Hooft-Feynman gauge
we add extra degrees of freedoms, the ghosts fields, in order to cancel the unphysical
degrees of freedom. We found a very good agreement in both gauges for the 
development of the zero mode and the energy. The influence of the various degrees
of freedoms which arise in gauge theories have led to new and interesting
results in the behavior of the zero mode. Due to the possibility for the Higgs field
to decay into other fields, the zero mode was efficiently damped. This behavior 
was until now not observed for systems treated in the one loop approximation beside for
the case of fermion decay \cite{Baacke:1998di}. As in the one loop approximation
in the $\phi^4$~theory the evolution of the zero mode in the unstable region was impossible.

For future studies there are two different interesting aspects. The first one is more technical.
In order to enter the unstable regime of the effective potential it is necessary
to go beyond the one loop approximation. Therefore, the implementation of Hartree-like
approximation would be very interesting. As we have shown here one has to include
higher order effects in a careful way in order to avoid IR-problems. 
In addition one has to choose very carefully an expansion parameter in order to secure
gauge invariance. Since the usual Hartree-approximation is just an ansatz and
an inconsistent summation and not an expansion in the coupling this will probably
lead to problems. Also the
investigation of the non-abelian gauge would be very interesting. 
Beside these technical considerations, the implementation of our
results for the gauge fields in a cosmological context would be very interesting.
Together with our studies on fermionic systems out of equilibrium we now
have built the foundation to examine more realistic models to
 describe the physics of the early universe. 
The recent success in detecting
neutrino masses has revived the idea of grand unification. An implementation
 of our method in Grand Unified Theories could lead to new and interesting results.
Also the implementation of Friedman-Robertson-Walker cosmology in the
model we have considered is interesting in order to get a more suitable model
for describing the inflationary scenario.


\acknowledgments
I would like to thank J. Baacke for motivating this work. He as influenced
this paper very much due to many fruitful discussions and ideas.  
I would also like to thank D. Cormier and E. Mottola for useful
discussions.

\begin{appendix}
\section{Perturbative Expansion}
\label{appA}
In this appendix we give some details of the perturbative expansion
of the mode functions which we have used for the renormalization procedure.
We have developed this formalism for non-equilibrium systems in
\cite{Baacke:1997se} and extended it for gauge theories in \cite{Baacke:1997kj}.
For more details the reader is referred to these works. First of all we have to
quantize the Goldstone field $\varphi$ and introduce the mode representation 
in the same way as the transversal 
gauge mode and the Higgs mode in Eqs. (\ref{trans}) and (\ref{isosc}).
Therefore, we have to investigate the following equation for the Goldstone field
(\ref{coulch})
\be\label{coulm}
\ddot\varphi+\omega_{e\lambda}^2(t)\varphi
-\frac{2e^2\dot \phi}{\omega_a^2}
\left[\phi\dot\varphi-\dot \phi\varphi\right]=0\pkt
\ee
In order to quantize the field we have 
to find the conjugate momentum. We can read it of from the Lagrangian
(\ref{lagr}):
\be
\Pi_\varphi=\dot\varphi-e a_0\phi
=\dot\varphi\frac{k^2}{\omega_a^2}
-\frac{e^2\phi\dot\phi}{\omega_a^2}\varphi\pkt
\ee
The commutation relation for the field and its momentum is given by
\be
[\varphi,\Pi_\varphi]=i\delta (\vec x-\vec x')\pkt
\ee
By computation of the commutator for the field and its time derivative,
we find that it is multiplied
by a factor
\be
[\varphi,\dot\varphi]\frac{k^2}{\omega_a^2}=i\delta(\vec x-\vec x')\pkt
\ee
Now we can expand the field in terms of the mode functions and 
the corresponding annihilation and creation operators
\be
\varphi=\intk\frac{1}{2\omega}\left(a_kU_\varphi(t)e^{i\vec k\cdot\vec x}
+a_k^\dagger U_\varphi^*(t)e^{-i\vec k\cdot\vec x}\right)\pkt
\ee
In order to satisfy the commutator for $\varphi$ and $\dot\varphi$,
the commutator for $a_k$ and $a_k^\dagger$ differs from its common form
\be
[a_k,a_k^\dagger]=2i(2\pi)^3\delta^3(\vec k-\vec k')\frac{1}{k^2}\pkt
\ee
The expectation value for the field is then given by
\be
\langle\varphi\varphi\rangle=\intk\frac{|U_\varphi|^2}{2\omega k^2}\pkt
\ee
The mode function satisfies the differential equation
\be
\ddot U_\varphi+\omega_{e\lambda}^2U_\varphi-\frac{2e^2\dot \phi}{\omega_a^2}
\left(\phi\dot U_\varphi-\dot \phi U_\varphi\right)=0\pkt
\ee
The frequency in the expectation value has to be fixed by the determination
of the Wronskian which belongs to the mode equation. Since the form of
the differential equation is not standard due to the first derivative of $U_\varphi$,
 the following transformation is efficient
\be
U_\varphi=\omega_a\tilde U_\varphi\pkt
\ee
The $\dot U_\varphi$-terms are canceled and we find a new mode
equation of the form
\be
\ddot{\tilde U_\varphi}+\left(\omega_{e\lambda}^2+
\frac{3e^2\dot\phi^2 k^2}{\omega_a^4}+\frac{e^2\phi\ddot\phi}{\omega_a^2}
\right)\tilde U_\varphi=0\pkt
\ee
Now we redefine the frequency $\omega^2_{e\lambda}$ in order to find a suitable
Wronskian. We choose
\be
\tilde \omega_{e\lambda}^2=\omega_{e\lambda}^2
+\frac{3e^2\dot\phi^2k^2}{\omega_a^4}+\frac{e^2\phi\ddot\phi}{\omega_a^2}\kma
\ee
and get the new differential equation for the mode function
\be
\ddot{\tilde U_\varphi}+\tilde\omega_{e\lambda}^2\tilde U_\varphi=0\pkt
\ee
The Wronskian then has the well known form
\be
\dot{\tilde U_\varphi}\tilde U_\varphi^*
-\tilde U_\varphi\dot{\tilde U}_\varphi^*=C\pkt
\ee
Since $\tilde U_\varphi$ behaves like $e^{-i\tilde\omega_{e\lambda 0}t}$
we can fix the constant to be $C=-2i\tilde\omega_{e\lambda 0}$. For the initial
time, $\dot\phi$ and the fluctuation integral vanish, so that 
$\tilde\omega_{e\lambda 0}^2$ simplifies to
\be
\tilde\omega^2_{e\lambda 0}=k^2+e^2\phi_0^2+\lambda(\phi^2_0-v^2)
-\frac{e^2\lambda}{\omega_{a0}^2}\phi_0^2\left(\phi_0^2-v^2\right)\pkt
\ee
For high momenta, which are important for the UV-divergences, the last
term is negligible. 

Now all mode equations have the similar structure
\bea
\left[\frac{d^2}{dt^2}+\omega_{j0}^2\right]U_j(t)&=&-V_j(t)U_j(t)\kma\\
\left[\frac{d^2}{dt^2}+\tilde\omega_{e\lambda 0}^2\right]\tilde U_\varphi(t)
&=&-\tilde V_{e\lambda}(t)\tilde U_\varphi(t)\kma
\eea
with $j=a,h$. Here we have introduced the potentials
\bea
V_a(t)&=&e^2[\phi^2(t)-\phi^2_0]\kma\\
V_{h}(t)&=&3\lambda\left[\phi^2(t)-\phi^2_0\right]\kma\\
\tilde V_{e\lambda}(t)&=&V_{e\lambda}(t)
+\frac{3e^2\dot \phi(t)^2k^2}{\omega_a^4(t)}
+\frac{e^2\phi(t)\ddot \phi(t)}{\omega_a^2(t)}
+\frac{e^2\lambda\phi_0^2(\phi_0^2-v^2)}{\omega_{a0}^2}\kma\\
V_{e\lambda}&=&\left(e^2+\lambda\right)\left(\phi^2-\phi_0^2\right)\pkt
\eea
We separate the mode functions into a trivial part corresponding to the case
that the potential vanishes and a function $f_j(t)$ which represents the reaction to the
potential by making the ansatz
\bea
\label{ansatzj}
U_j(t)&=&e^{-i\omega_{j0}t}\left[1+f_j(t)\right]\kma\\
\tilde U_\varphi(t)&=&e^{-i\tilde\omega_{e\lambda 0} t}\left[1+f_\varphi(t)\right]\pkt
\eea
Then the new mode functions satisfy the differential equation
\be
\ddot f_j(t)-2i\omega_{j0}\dot f_j(t)=-V_j(t)[1+f_j(t)]\kma
\ee
respectively for $f_\varphi$
with the initial conditions $f_j(0)=\dot f_j(0)=0$. Expanding now
$f_j(t)$ with respect to orders in the potential by writing
\be
f_j(t)=f_j^{(1)}+f_j^{(2)}+f_j^{(3)}+\cdots\kma
\ee
we can extract the leading behavior for $f_j(t)$:
\bea
\label{lofj}
f_j^{(1)}(t)&=&-\frac{i}{2\omega_{j0}}\inttt V_j(t')
-\frac{V_j(t)}{4\omega_{j0}^2}
+\frac{1}{4\omega_{j0}^2}\inttt 
e^{2i\omega_{j0}\Delta t}\dot V_j(t')+{\cal O}(\omega^{-3}_{j0})\kma\\
f_j^{(2)}(t)&=&-\frac{1}{4\omega_{j0}^2}\inttt\int\limits_{0}^{t'}\!{d}t''\,
V_j(t')V_j(t'')+{\cal O}(\omega_{j0}^{-3})\kma
\eea
with $\Delta t=t-t'$. For $f_\varphi$ the frequency and the potential have to 
be replaced by $\tilde\omega_{e\lambda 0}$ and $\tilde V_{e\lambda}$.

Inserting the leading behavior of the new mode functions into the 
fluctuation integral (\ref{ffin}) allows us to extract the following divergences
\bea
\calf_{\rm div}&=&
3\lambda\phi\intk\frac{1}{2\omega_{h0}}
\left(1-\frac{V_h}{2\omega_{h0}}\right)\nonumber\\
&&+2e^2\phi\intk\frac{1}{2\omega_{a0}}
\left(1-\frac{V_a}{2\omega_{a0}}\right)\nonumber\\
&&+\lambda\phi\intk\frac{1}{2\tilde\omega_{e\lambda 0}}
\left(1+\frac{e^2\phi^2}{k^2}-\frac{V_{e\lambda}}{2\tilde\omel^2}\right)
\nonumber\\
&&+e^2\phi\intk\frac{1}{2\tilde\omega_{e\lambda 0}}
\left(1+\frac{e^2m_\varphi^2}{k^2}-\frac{V_{e\lambda}}{2\tilde\omel^2}
\right)\pkt
\eea
Dimensional regularization leads to the counter terms given in Eqs.
(\ref{ctm})-(\ref{ctz}) and finite corrections.

In a similar way we have to treat the energy density (\ref{enmod}).
Here we find with the help of the truncated mode functions $f$
\bea
\cale_{\rm div}&=&
\intk\left(\frac{\omega_{h0}}{2}+\frac{V_h}{4\omega_{h0}}-\frac{V_h^2}{16\omega_{h0}^3}\right)
\nonumber\\
&&+\intk\left(\omega_{a0}+\frac{V_a}{2\omega_{a0}}-\frac{V_a^2}{8\omega_{a0}^3}\right)
\nonumber\\
&&-e^2\dot\phi^2\intk\frac{1}{4k^2\tilde\omega_{e\lambda 0}}\\
&&+\intk\left[\frac{\tilde\omega_{e\lambda 0}}{2}
+\frac{V_{e\lambda}}{4\tilde\omega_{e\lambda 0}}
+\frac{m_a^2m_\varphi^2}{4k^2\tilde\omega_{e\lambda 0}}
-\frac{V_{e\lambda}^2}{16\tilde\omega_{e\lambda 0}^3}
+\frac{m_{\varphi 0}^2m_{W0}^2}{4\oma^2\tilde\omel}
\right]\nonumber
\eea
which leads again to $\delta m$, $\delta\lambda$, and $\delta Z$
and in addition to $\delta \Lambda$ given in Eq. (\ref{cosm}) and also
to an extra finite contribution.


\section{The leading Feynman diagrams}
\label{appB}
In order to understand the behavior of the system it is very instructive to
analyse the leading Feynman diagrams in detail. This is especially
easy in the 't Hooft-Feynman gauge, since we can read of the Feynman rules
directly from the gauge fixed Lagrangian, which is given for example in
\cite{Baacke:1997kj}. We find four different propagators which are
\begin{enumerate}
\item the gauge boson propagator:
\begin{equation}
\parbox{4,5cm}
{\mbox{\epsfxsize=4cm\epsfbox{eich.eps}}}
\begin{array}{c}
i\Delta^{ab}_{\mu\nu}=-\frac{ig_{\mu\nu}\delta^{ab}}{k^2-m_W^2+i\epsilon}
\kma\\[4mm]
\mbox{with } m_W^2=e^2 v^2\kma\\
\end{array}
\end{equation}
\item the propagator for the isoscalar Higgs field:
\begin{equation}
\parbox{4,5cm}
{\mbox{\epsfxsize=4cm\epsfbox{h.eps}}}
\begin{array}{c}
i\Delta_{h}=i\frac{1}{k^2-m_h^2+i\epsilon}\kma\\[4mm]
\mbox{with } m_h^2=2\lambda v^2.\\
\end{array}
\end{equation}
\item the propagator for the Goldstone field:
\begin{equation}
\parbox{4,5cm}
{\mbox{\epsfxsize=4cm\epsfbox{phi.eps}}}
\begin{array}{c}
i\Delta^{ab}_\varphi=i\frac{\delta^{ab}}{k^2-m_W^2+i\epsilon}\kma\\[4mm]
\end{array}
\end{equation}
\item the propagator for the ghost field:
\begin{equation}
\parbox{4,5cm}
{\mbox{\epsfxsize=4cm\epsfbox{geist.eps}}}
\begin{array}{c}
i\Delta^{ab}_\eta=i\frac{\delta^{ab}}{k^2-m_W^2+i\epsilon}\kma\\[4mm]
\end{array}
\end{equation}
\end{enumerate}
and ten vertices which are relevant if we restrict ourselves to second
order in the fluctuations

\vspace{0.5cm}

\parbox{15cm}{
\parbox[t]{7cm}
{V1)\begin{center}
\mbox{\epsfxsize=2.5cm\epsfbox{g1n.eps}}
\end{center}}
\hspace{.5cm}\parbox[t]{7cm}
{V2)\begin{center}
\mbox{\epsfxsize=2.5cm\epsfbox{g3.eps}}
\end{center}}

\parbox[t]{7cm}
{\begin{center}
$i\Gamma=-i\frac{3}{2}\lambda\left[\phi^2(t)-v^2\right]$
\end{center}}
\hspace{.5cm}\parbox[t]{7cm}
{\begin{center}
$i\Gamma=-i\left(\frac{\lambda}{2}+\frac{e^2}{2}\right)\delta^{ab}\left[
\phi^2(t)-v^2\right]$
\end{center}}}

\parbox{15cm}{
\parbox[t]{7cm}
{V3)\begin{center}
\mbox{\epsfxsize=2.5cm\epsfbox{g2n.eps}}
\end{center}}
\hspace{.5cm}\parbox[t]{7cm}
{V4)\begin{center}
\mbox{\epsfxsize=2.5cm\epsfbox{g4n.eps}}
\end{center}}

\parbox[t]{7cm}
{\begin{center}
$i\Gamma=i\frac{e^2}{2}\delta^{ab}g_{\mu\nu}\left[\phi^2(t)-v^2\right]$
\end{center}}
\hspace{.5cm}\parbox[t]{7cm}
{\begin{center}
$i\Gamma=-i {e^2}\delta^{ab}\left[\phi^2(t)-v^2\right]$
\end{center}}}

\parbox{15cm}{
\parbox[t]{7cm}
{V5)\begin{center}
\mbox{\epsfxsize=2.5cm\epsfbox{g5n.eps}}
\end{center}}
\hspace{.5cm}\parbox[t]{7cm}
{V6)\begin{center}
\mbox{\epsfxsize=2.5cm\epsfbox{g16n.eps}}
\end{center}}

\parbox[t]{7cm}
{\begin{center}
$i\Gamma=-i\lambda \phi(t)$
\end{center}}
\hspace{.5cm}\parbox[t]{7cm}
{\begin{center}
$i\Gamma=-i\delta^{ab} \phi(t)\left(\lambda+ e^2\right)$
\end{center}}}

\parbox{15cm}{
\parbox[t]{7cm}
{V7)\begin{center}
\mbox{\epsfxsize=2.5cm\epsfbox{g17n.eps}}
\end{center}}
\hspace{.5cm}\parbox[t]{7cm}
{V8)\begin{center}
\mbox{\epsfxsize=2.5cm\epsfbox{g6n.eps}}
\end{center}}

\parbox[t]{7cm}
{\begin{center}
$i\Gamma=i e^2\delta^{ab}g_{\mu\nu}\phi(t)$
\end{center}}
\hspace{.5cm}\parbox[t]{7cm}
{\begin{center}
$i\Gamma=-2i e\delta^{ab}\phi(t)$
\end{center}}}

\parbox{15cm}{
\parbox[t]{7cm}
{V9)\begin{center}
\mbox{\epsfxsize=2.5cm\epsfbox{g14.eps}}
\end{center}}
\hspace{.5cm}\parbox[t]{7cm}
{V10)\begin{center}
\mbox{\epsfxsize=2.5cm\epsfbox{g13n.eps}}
\end{center}}

\parbox[t]{7cm}
{\begin{center}
$i\Gamma=i e^2\epsilon^{abc}q_\mu\phi(t)$
\end{center}}
\hspace{.5cm}\parbox[t]{7cm}
{\begin{center}
$i\Gamma=2i e\delta^{ab}q_\mu  \phi(t)$
\end{center}}}

These lead to the following Feynman graphs up to second order in perturbation theory

\vspace{0.5cm}
\begin{equation}
\parbox{4,5cm}
{\mbox{\epsfxsize=3cm\epsfbox{graphn.eps}}}
\end{equation}

\vspace{0.5cm}

\parbox{15cm}{
\parbox[t]{7cm}
{\begin{center}
\mbox{\epsfxsize=3cm\epsfbox{graphe.eps}}
\end{center}}
\hspace{.5cm}\parbox[t]{7cm}
{\begin{center}
\mbox{\epsfxsize=3cm\epsfbox{phie.eps}}
\end{center}}
\parbox[t]{7cm}
{\begin{center}
\mbox{\epsfxsize=3cm\epsfbox{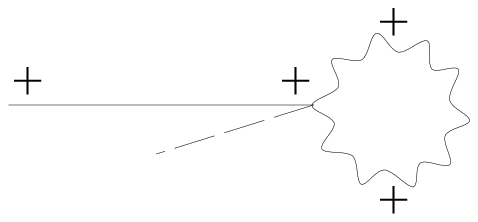}}
\end{center}}
\hspace{.5cm}\parbox[t]{7cm}
{\begin{center}
\mbox{\epsfxsize=3cm\epsfbox{geiste.eps}}
\end{center}}}

\parbox{15cm}{
\parbox[t]{7cm}
{\begin{center}
\mbox{\epsfxsize=3cm\epsfbox{graphz.eps}}
\end{center}}
\hspace{.5cm}\parbox[t]{7cm}
{\begin{center}
\mbox{\epsfxsize=3cm\epsfbox{phiz.eps}}
\end{center}}
\parbox[t]{7cm}
{\begin{center}
\mbox{\epsfxsize=3cm\epsfbox{az.eps}}
\end{center}}
\hspace{.5cm}\parbox[t]{7cm}
{\begin{center}
\mbox{\epsfxsize=3cm\epsfbox{geistz.eps}}
\end{center}}}

\parbox{15cm}{
\parbox[t]{7cm}
{\begin{center}
\mbox{\epsfxsize=3cm\epsfbox{gekkb.eps}}
\end{center}}}

\begin{figure}[tb]

\vspace{-2.3cm}

\begin{picture}(13,10)
\put(0,0){\epsfxsize=13cm\epsfbox{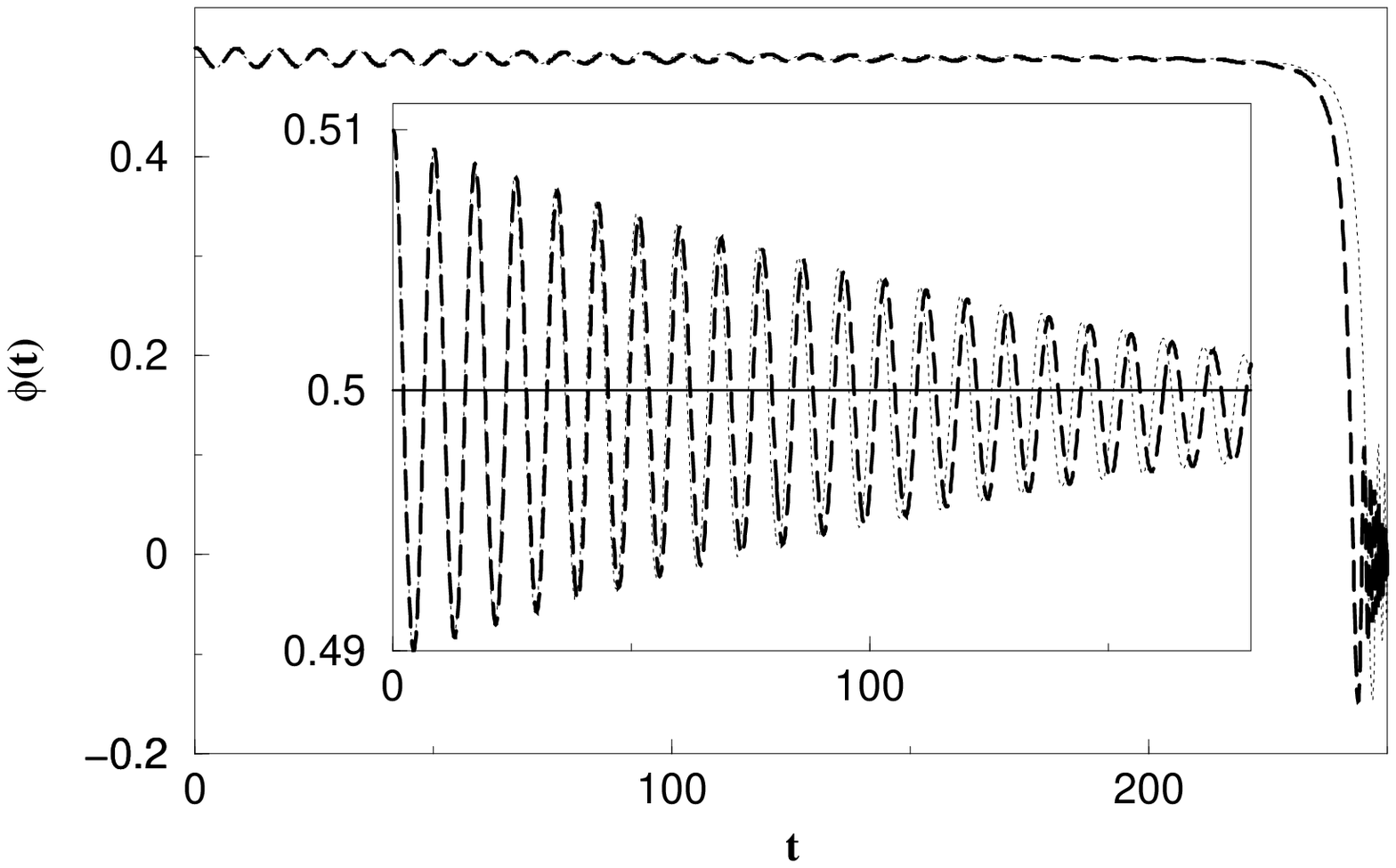}}
\end{picture}
\caption{Zero mode vs. $t$ for parameter set 3, dashed line: Feynman gauge, dotted line: Coulomb gauge;
frame: zoom onto a shorter time scale, solid line: critical value for $\phi(t)$.}
\label{inst}

\vspace{-1.5cm}

\begin{picture}(13,10)
\put(0,0){\epsfxsize=13cm\epsfbox{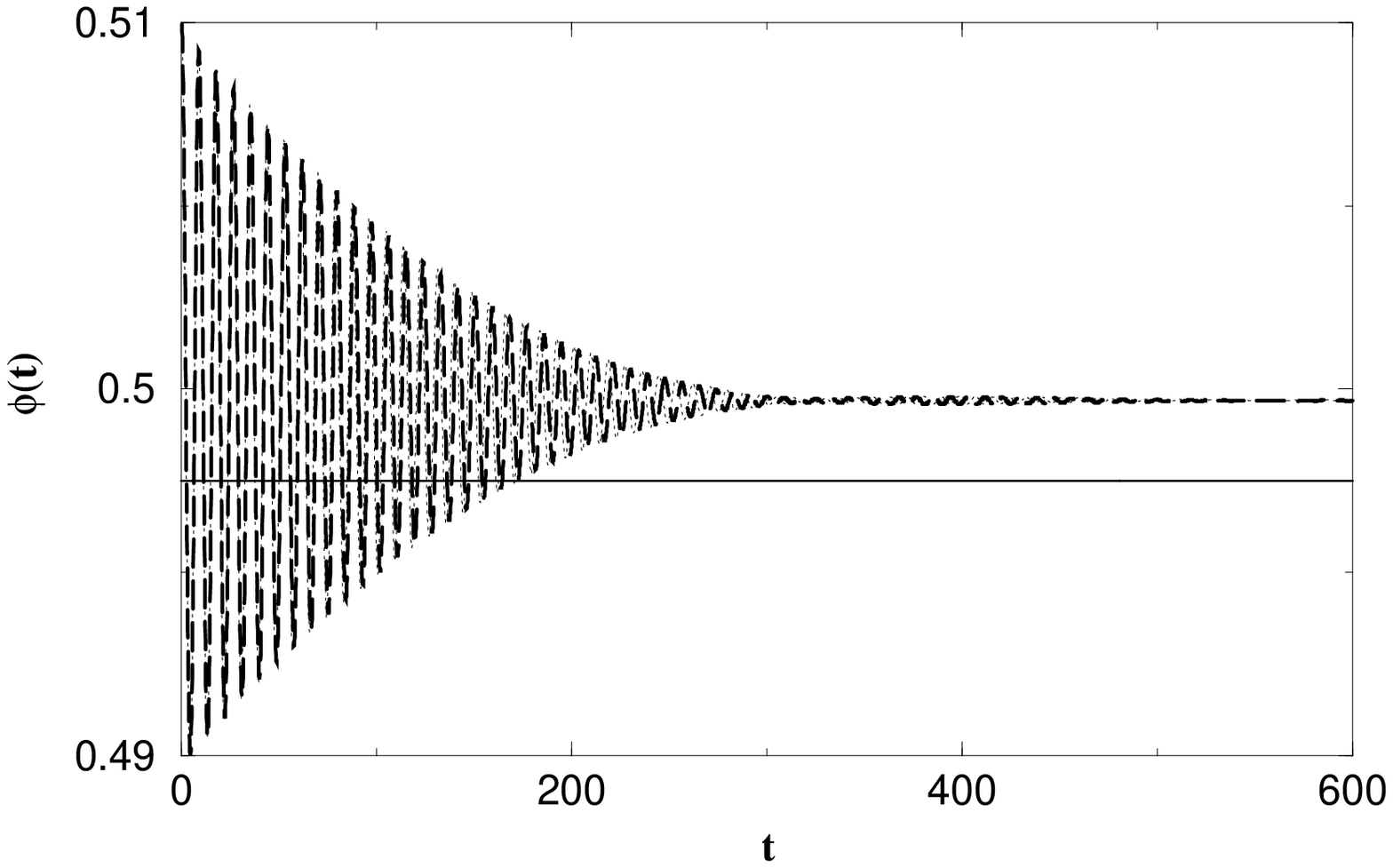}}
\end{picture}
\caption{Zero mode vs. $t$ for parameter set 4, dashed line: Feynman gauge, dotted line: Coulomb gauge,
solid line: critical value for $\phi(t)$.}
\label{l1e01}
\end{figure}

The plus and minus signs at the graphs indicates the use of propagators in the
CTP-formalism.
Especially interesting is the last graph; it leads to the possibility
of the decay of $h$ into $a_0$ and $\varphi$ which causes the strong damping
of the zero mode for parameter set 4 displayed in Fig. \ref{l1e01}.
The various degrees of freedom in gauge fields lead to new and interesting
results in non-equilibrium quantum field theory. A detailed analysis
of these graphs within the CTP-formalism is given in \cite{Heitmann:2000}.
\end{appendix}

\clearpage

\begin{table}
\begin{center}
\begin{tabular}{|c|c|c|c|c|c|c|c|}
\hline
\hfill &$\lambda$&$e=\frac g 2$&$m_h^2$&$\phi_0$&
$m^2_{a0}$&$m^2_{h0}$&$m_{e\lambda0}^2$\\
\hline	
Parameter set 1&1&1&0.5&0.51&0.26&0.53&0.27\\
Parameter set 2&0.33&1.3&0.5&0.51&0.44&0.01&0.28\\
Parameter set 3&1&0.001&0.5&0.51&2.6$\cdot 10^{-7}$&0.53&0.01\\
Parameter set 4&1&0.1&0.5&0.51&2.6$\cdot 10^{-3}$&0.53&0.01\\
\hline
\end{tabular}
\caption{Parameter sets}
\label{tab4}
\end{center}
\end{table}

\end{document}